\documentclass[aps,prl,twocolumn,groupedaddress]{revtex4-2}

\usepackage{pgfplots}
\newenvironment{customlegend}[1][]{%
    \begingroup
    \csname pgfplots@init@cleared@structures\endcsname
    \pgfplotsset{#1}%
}{%
    \csname pgfplots@createlegend\endcsname
    \endgroup
}%
\def\addlegendimage{\csname pgfplots@addlegendimage\endcsname}

\newcommand{\addlegendimageintext}[1]{%
    \tikz {
        \begin{customlegend}[
            legend entries={\empty},
            legend style={
                draw=none,
                inner sep=0pt,
                column sep=0pt,
                nodes={inner sep=0pt}}]
        \addlegendimage{#1}
        \end{customlegend}
    }%
}
\begin{document}


\title{Nanowire networks: how does small-world character evolve with dimensionality?}


\author{Ryan K. Daniels}
\affiliation{The MacDiarmid Institute for Advanced Materials and Nanotechnology, School of Physical and Chemical Sciences, University of Canterbury, Christchurch, New Zealand}

\author{Simon A. Brown}
\affiliation{The MacDiarmid Institute for Advanced Materials and Nanotechnology, School of Physical and Chemical Sciences, University of Canterbury, Christchurch, New Zealand}


\date{\today}

\begin{abstract}
    Networks of nanowires are currently under consideration for a wide range of electronic and optoelectronic applications. Nanowire devices are usually made by sequential deposition, which inevitably leads to stacking of the wires on top of one another. Here we demonstrate the effect of stacking on the topology of the resulting networks. We compare perfectly 2D networks with quasi-3D networks, and compare both nanowire networks to the corresponding Watts Strogatz networks, which are standard benchmark systems. By investigating quantities such as clustering, path length, modularity, and small world propensity we show that the connectivity of the quasi-3D networks is significantly different to that of the 2D networks, a result which may have important implications for applications of nanowire networks.
\end{abstract}


\maketitle

\section{Introduction}

Networks of nanowires and nanotubes have a range of potential applications in photovoltaic, optoelectronic, and sensor devices such as flexible transparent electrodes, organic light emitting diodes, and bioelectric interfaces.\cite{Ye2014, Park2013, Sannicolo2016, Tian2019, Fennell2016, Zheng2016, Song2016, Lee2017, Li2020} More particularly, it has been proposed that the structures of self-assembled networks of nanowires (and nanoparticles\cite{Mallinson2019, Shirai2020, Pike2020}) mimic some  of the complex biological structures of the brain, and therefore may be useful as novel systems for neuromorphic computing.\cite{Avizienis2012,Manning2018, Tanaka2018}  Networks of nanowires exhibit a number of brain-like properties such as reconfiguration dynamics and short and long term memory.\cite{Stieg2012,Diaz2019,Milano2020} They exhibit winner-takes-all behaviour\cite{Manning2018} and have been proposed as  systems for physical reservoir computing,\cite{Avizienis2012} able to perform tasks such as nonlinear transformations, sine wave auto-generation, and Mackey-Glass chaotic time series forecasting.\cite{Fu2020, Kuncic2020}

Randomly deposited networks of nanowires have been studied extensively within the framework of percolation theory.\cite{Stauffer1992, Fostner2014,Li2009,Langley2018,White2010, Mutiso2013,Loeffler2020} More recent work has  addressed the topological structure of 2D nanowire networks using well-established techniques in graph and network theory. These networks have been shown to exhibit a small-world architecture similar to many biological systems,\cite{Loeffler2020, Pantone2018} i.e. the nodes are considered to be both highly clustered and have small path lengths between nodes.\cite{Watts1998} This concept is of interest because small-world architecture is thought to affect emergent behavior in complex systems,\cite{Pacual2016, Strogatz2001} and to be important for information propagation across  networks,\cite{Nishikawa2003, Lu2004}  with several groups also demonstrating that reservoir computing performance is impacted by  small-world or scale-free network architectures.\cite{Haluszczynski2020, Deng2007} 

Previous modelling  has however assumed that the nanowire networks are perfectly two-dimensional i.e. that the  wires lie in a plane.\cite{Li2009,Langley2018,White2010, Mutiso2013,Loeffler2020, Pantone2018} This is clearly not the case for real-world nanowire networks in which the wires are stacked on top of one another during a deposition process. Given the importance of topological structure for information processing and a range of other network properties  there is clearly a need for models which incorporate realistic stacking of the nanowires.
\begin{figure*}[ht]
	\centering
	\includegraphics[height=7.3cm]{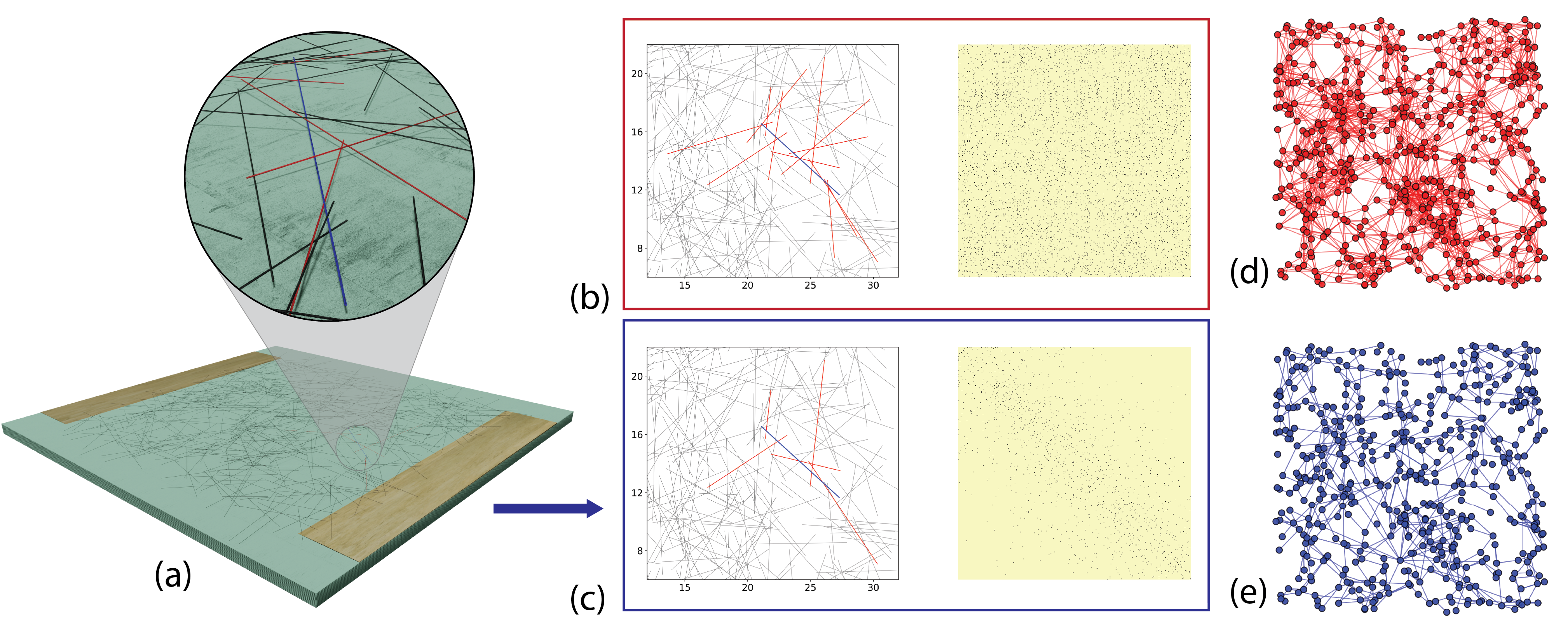}
	\caption{\textit{(a) Schematic illustration of a simple device with two contacts. The network of 500 nanowires is deposited on a $30\mu$m$\times30\mu$m drop zone. The length of each wire was drawn from a uniform distribution with mean length $6 \mu$m and range $[3,9]\mu$m. The magnified part of (a) shows the  stacking of the wires in the vertical direction after sequential deposition. The blue wire is a randomly chosen wire which makes connections with other wires (in red) during the deposition process. (b), (c) An example of the difference in connectivity between the two networks when stacking is included in the model. Left: a portion of the network; right: the adjacency matrix. In the 2D network (b), the chosen wire forms a junction with every wire that it intercepts (red), leading to many more connections. However in the Q3D network (c) the chosen wire connects only to the limited number of wires that are in direct contact.  (d, e) The graph representation of the same networks. The Q3D network  has fewer edges than the 2D network and thus has  a much sparser adjacency matrix than the 2D network, as shown in (b) and (c).}}
	\label{fgr:schematic}
\end{figure*}

In the present work, we seek to understand the impact of stacking on the connectivity of a nanowire network. We build a physical model that explicitly considers the sequence in which the wires are deposited and the consequent stacking of the wires in a three-dimensional volume. It is important to note that because the network is built up by sequential deposition, it is  \emph{quasi}-three-dimensional (Q3D) and is therefore distinct from percolating systems\cite{Stauffer1992} in which the objects are randomly placed in a 3D space (without requiring that the wires rest on each other).
We show here that perfectly 2D networks have much broader degree distributions and much larger mean degrees than the Q3D networks. We demonstrate the impact of dimensionality on the small-world nature of the networks, and conclude that experimentally realized Q3D networks of nanowires are likely to have very different network structures than model 2D structures.

\section{Nanowire Models}
We consider networks of $N$ wires,  deposited on a $\Lambda\times\Lambda$ $\mu$m area, which is considered to be the $xy$-plane. For each wire in the network, the $x$- and $y$-coordinates of the mid-point of the wire were chosen at random from a uniform distribution over $[0, \Lambda]$, and the angle the wire makes with the $x$-axis was drawn from a uniform distribution over $[-\pi/2, \pi/2]$. Every point where two or more wires touch is taken to be a junction in the network. This connectivity is then stored in a $N\times N$ adjacency matrix, which also allows the number of connections between each wire and its neighbours (the degree, $k$) to be calculated. For the 2D network, every wire is considered to be a 1D line (i.e. with zero thickness). As a result, the wires lie exactly in the $xy$-plane, and every intercept is a junction (similar to that shown in Fig. \ref{fgr:schematic}c). 

In the Q3D network, each wire also has a diameter and therefore carves out a volume in space that cannot be occupied by any other wire. The wires are deposited sequentially, one at a time, and the vertical position of each new wire depends on the positions of the wires that have already been deposited. A comparison of Figs. \ref{fgr:schematic}b and c shows that stacking causes some wires that are in contact in the perfectly 2D system to become separated vertically in the Q3D system. 

For each new wire, the algorithm determines firstly which wires are below it, and calculates the intercept coordinates,  with the largest $z$-value as the first contact point. This point acts as a pivot  around which the new wire must rotate. The location of the centre of mass with respect to the pivot point determines the direction of rotation, and hence the location of the second point of contact. This results in a network that is stacked vertically in a third dimension, the $z$-axis, as shown in Fig. \ref{fgr:schematic}a and ESI Fig. S1 and S2.  As with the 2D model, every point of contact is a junction and the connections between wires are stored in an adjacency matrix (Fig. \ref{fgr:schematic}b and c). This allows construction of a graph representation of the network, where each node is taken to be the center of a wire and each edge represents a  contact between two wires (Fig. \ref{fgr:schematic}d and e).

It is of course possible to consider additional processes during the wire deposition, including allowing the wires to settle, slide or bend. However, the purpose of the present work is to investigate the effect of dimensionality on network connectivity (rather than  the detailed deposition dynamics) and so we focus on the simplest model that captures the essential features, i.e. the wires are perfectly rigid and do not move from their initial position after deposition.

In order to compare the nanowire networks with benchmark networks, we also generated four sets of Watts-Strogatz (WS) networks. The first step in generating WS networks is to calculate the  mean  degree $\bar{k}$ for each 2D or Q3D network. The WS network has the same $N$ and $\bar{k}$ but  connections are rewired with probability $p$ to another node chosen  at random from all other available nodes.\cite{Watts1998} As shown in ESI Fig S3, for each value of $\bar{k}_N$ we construct two WS networks: one with $p=0$, making a regular network, and the other with $p=1$, making a random network.  We then compare the 2D, Q3D, and the WS networks using a number of common measures from network theory (see  Methods). These include the degree, path length ($L$), and average local clustering coefficient ($C$) distributions. We then also examine the community structure of the network, the modularity of a community partitioned network, and the small world propensity. 

\section{Results and discussion}
\begin{figure}[h]
	\centering
	\includegraphics[height=6.7cm]{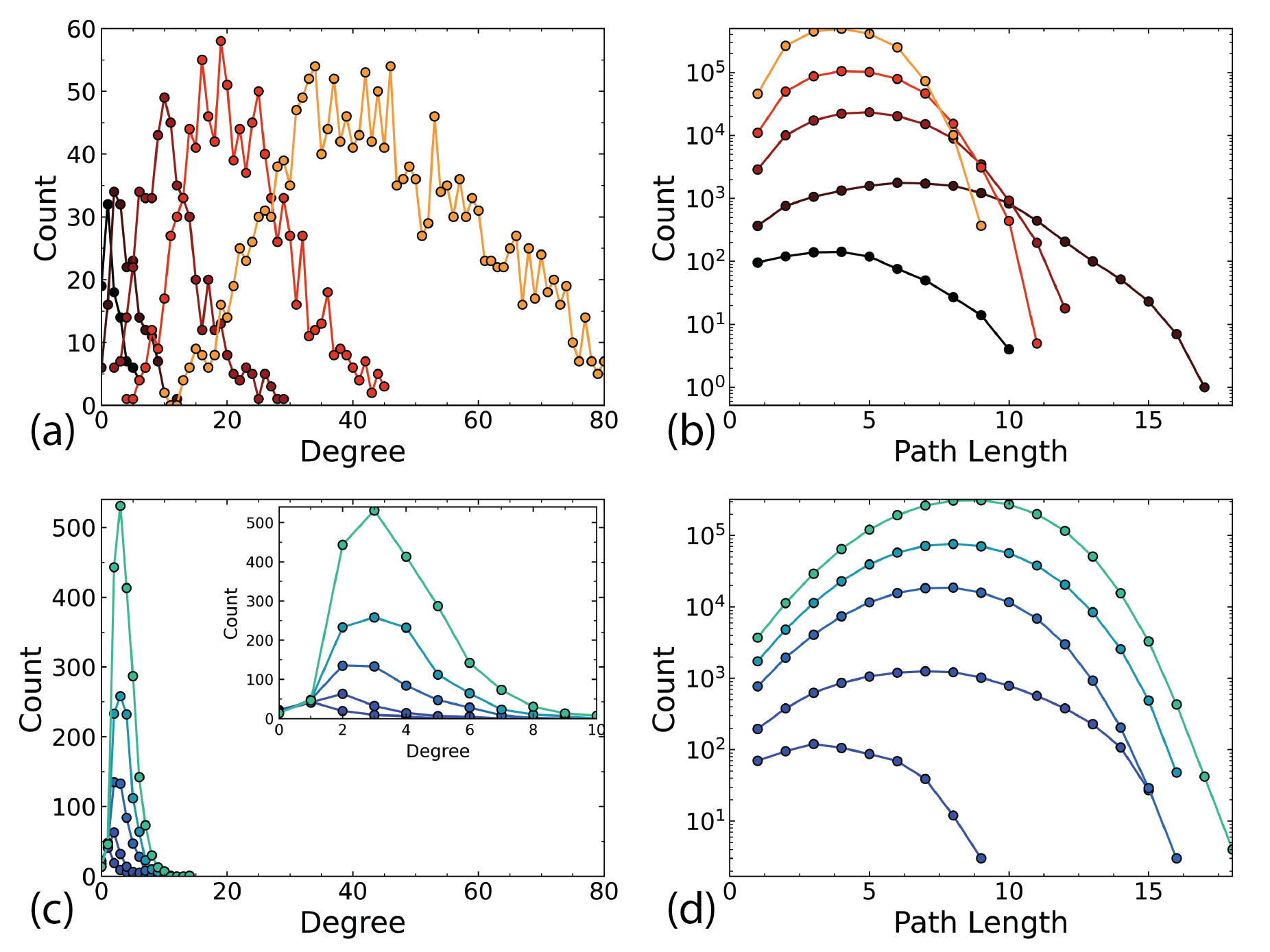}
	\caption{\textit{(a), (b) Degree distribution and path length for a 2D network with $N=100, 180, 500, 1000, 2000$ (from dark to light colours) and $\lambda=6\mu$m, $d=0.5$. (c), (d) Degree distribution and path length distribution for a Q3D network with the same parameters. Note that figures in the same column are plotted on the same $x$-scale. The inset image in (c) shows the degree distribution of the Q3D network on a zoomed $x$-scale. Note the logarithmic scale for (b), (d) which allows visualisation of the broadening with increasing $N$ of the path length distribution of the Q3D network.}}
	\label{fgr:path_degree}
\end{figure}
\begin{figure}[h]
	\centering
	\includegraphics[height=8.9cm]{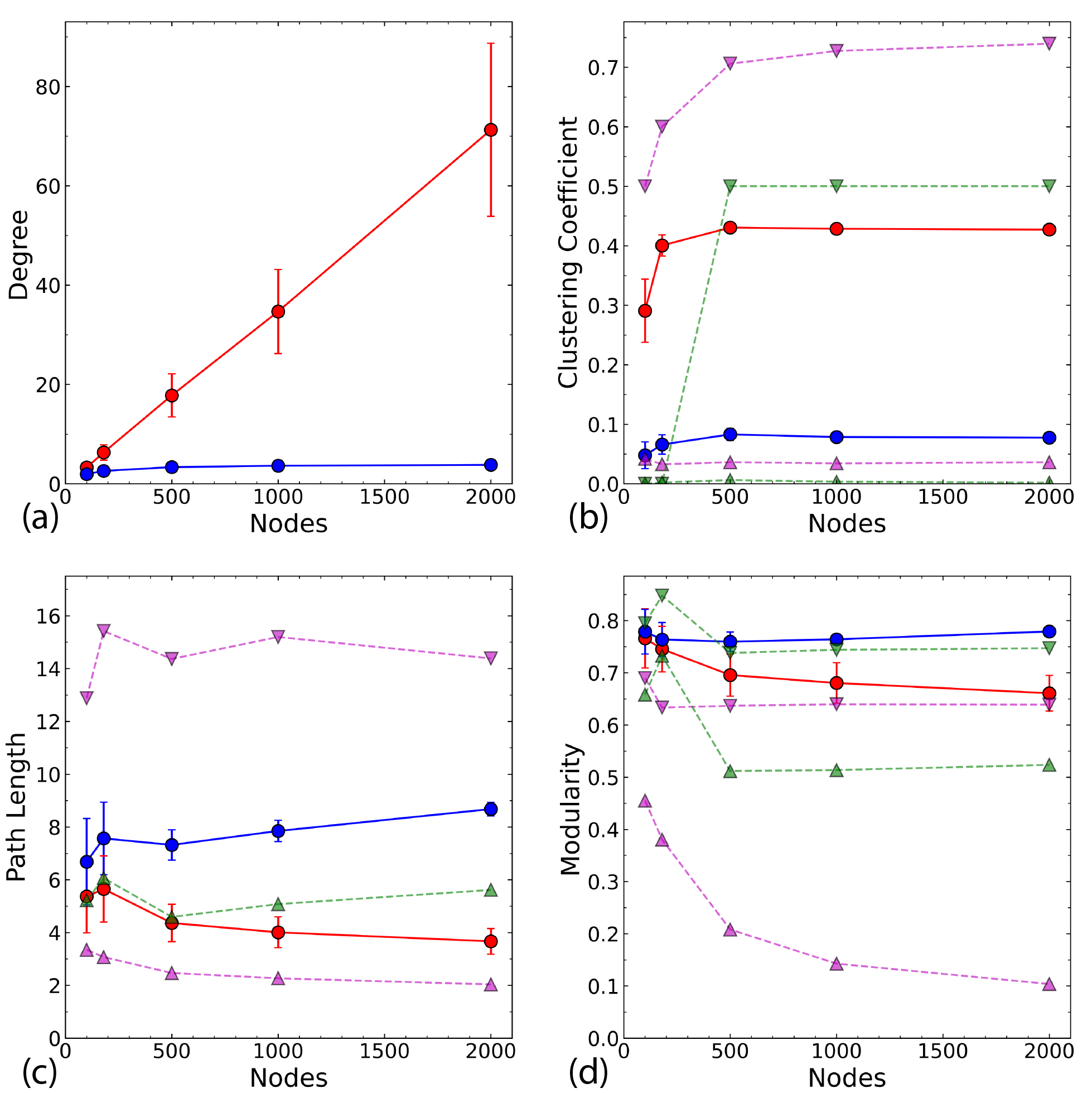}
	\caption{\textit{(a) The dependence of average degree $\bar{k}$ on system size. The markers are the average values taken over all combinations of $\lambda$ $(6, 6.5, 7, 7.5, 8, 8.5, 9)$ and $d$ $(0, 0.1, 0.2, 0.5)$ ($28$ networks) for both the 2D (red) and Q3D (blue) networks.  (b) Average clustering coefficient for the 2D and Q3D  networks with the corresponding Watts-Strogatz networks for comparison. For each value of the mean degree, two WS networks were constructed: one with a rewiring probability of $p=0$ (down triangle) and one with $p=1$ (up triangle). This was done for degree values of both the 2D (magenta) and the Q3D (green) networks. (c) The average path lengths of the network. (d) The resulting maximized modularity after the partitioning of the networks using the Louvain algorithm. All error bars are one standard deviation of the data obtained from all $28$ networks. Note that the regular WS ($p=0$) networks corresponding to the 3D network have high values of $L$, and so are not displayed here. A modified version of this plot with a logarithmic $y$-scale is presented in ESI Fig. S6.}}
	\label{fgr:4panel}
\end{figure}
We varied the mean wire length $\lambda$, the dispersion $d$, and the number of nanowires in the network $N$. Values of $\lambda$ were chosen to be 6, 6.5, 7, 7.5, 8, 8.5, and 9 $\mu$m. Values for the dispersion were chosen to be 0, 0.1, 0.2 and 0.5 (hence the minimum possible length of any given wire is $\lambda-d\lambda$ and the maximum possible length is $\lambda+d\lambda$). $\Lambda$ is fixed at $30 \mu$m and the scaling of the network properties with $\lambda$ and $\Lambda$ is shown in ESI Fig S5. These choices of parameters enables comparisons with existing literature on the topology of 2D nanowire networks.\cite{Loeffler2020} For the Q3D network, the wire diameter is chosen to be $20$ nm. We considered networks in which the number of wires ranged from 50 (almost perfectly two-dimensional) to 2000 (highly three-dimensional as shown in ESI Fig. S1 and S2). For clarity of presentation we show data only for $N =$ 100, 180, 500, 1000, and 2000 wires.

\subsection*{Degree and path length}

The difference in the connectivity of the networks (Fig. \ref{fgr:schematic}c, d) is reflected in the degree and path length distributions, as shown in  Fig. \ref{fgr:path_degree} for typical networks for each $N$. 
The 2D networks in Fig. \ref{fgr:path_degree}a, b have much broader degree distributions and generally narrower path length distributions than the Q3D networks (Fig. \ref{fgr:path_degree}c, d). In the 2D network the number of intersections increases for each newly deposited wire. Therefore, we see an approximately linear increase in mean degree with increasing N (Fig. \ref{fgr:4panel}a). We also see a gradual decrease in path length as more connections become available (Fig. \ref{fgr:4panel}c). In contrast, for the Q3D networks $\bar{k}$  is much lower  and it increases at a much slower rate. The growth of the height of the network (ESI Fig. S2) in the $z$-direction also results in a higher mean path length for the Q3D networks.
\begin{figure}[t]
	\centering 
	\includegraphics[height=5cm]{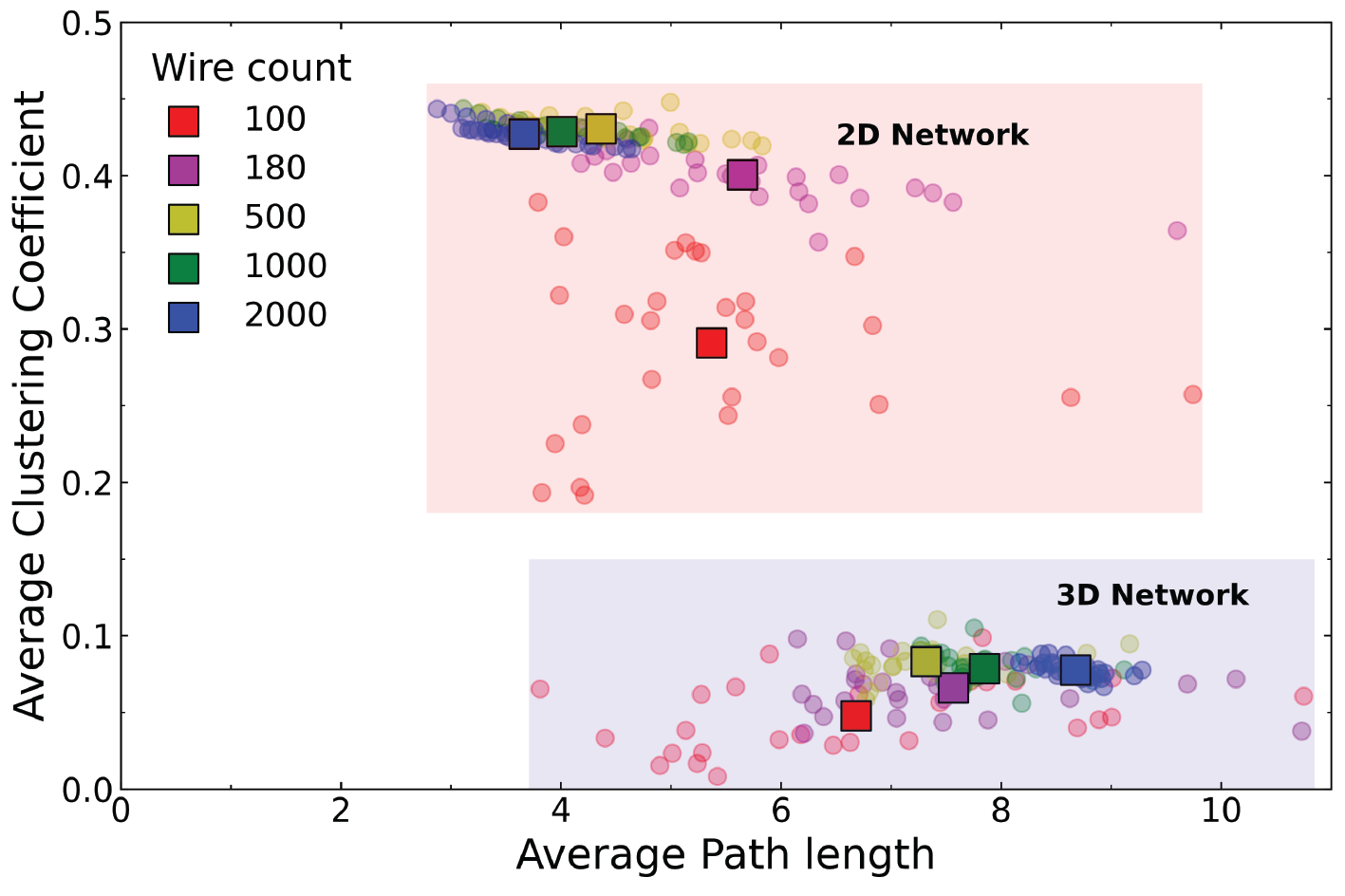}
	\caption{\textit{Watts-Strogatz cartographic plane showing the average clustering coefficient against  the average path length. The small circular markers are  results for each combination of $\lambda$ and $d$, with the different colours representing different $N$. The large squares are the mean values for each $N$. The data for the $N=180, 500, 1000,$ and $2000$ wire networks are closely grouped, but the data for  $N=100$  is widely dispersed due to the proximity of the percolation threshold (for both the 2D and Q3D systems $N_c \sim 50-150$).}}
	\label{fgr:carto}
\end{figure}
\subsection*{Clustering and comparison with WS networks}

Figs. \ref{fgr:4panel}b and c compare the clustering coefficients and path lengths of each nanowire network with the corresponding WS networks. The 2D network, has high clustering like the WS regular network ($p=0$, \addlegendimageintext{black, mark=triangle*, fill=magenta, mark size=3pt, only marks, rotate=180, opacity=0.7}),  but a low mean path length similar to that of the WS random network ($p=1$, \addlegendimageintext{black, mark=triangle*, fill=magenta, mark size=3pt, only marks, opacity=0.7}). These are typical characteristics of a small world network. The Q3D model, on the other hand, exhibits a low degree of clustering, similar to that of the corresponding WS random network (\addlegendimageintext{black, mark=triangle*, fill=black!40!green, mark size=3pt, only marks, opacity=0.7}). The mean path lengths in the Q3D network are higher than those in the 2D network, but are still similar to those in the corresponding WS random network.

Fig. \ref{fgr:carto} shows a comparison of the average mean path lengths and average local clustering coefficients for a large number of realisations of both the 2D and Q3D networks, with different values of $N$, $\lambda$, and $d$. This alternative representation of the data shows clearly that the Q3D networks are significantly less clustered and have greater mean path lengths than the 2D networks. The mean values of $L$ (square symbols) are similar for $N=100$ but become more different as $N$ increases: $\bar{L}$ decreases for the 2D networks, but increases for the Q3D networks. There is considerable scatter in the path lengths from network to network, but the clustering coefficients are only weakly dependent on $N$,  wire length and dispersion for $N$ > 100. For $N$=100, in both the 2D and Q3D networks, there is significant variation in clustering  because the system is close to the percolation threshold ($N_c \sim 50-150$, see ESI Fig. S4).

\subsection{Modularity and community structure}

\begin{figure}[t]
	\centering
	\includegraphics[height=4.5cm]{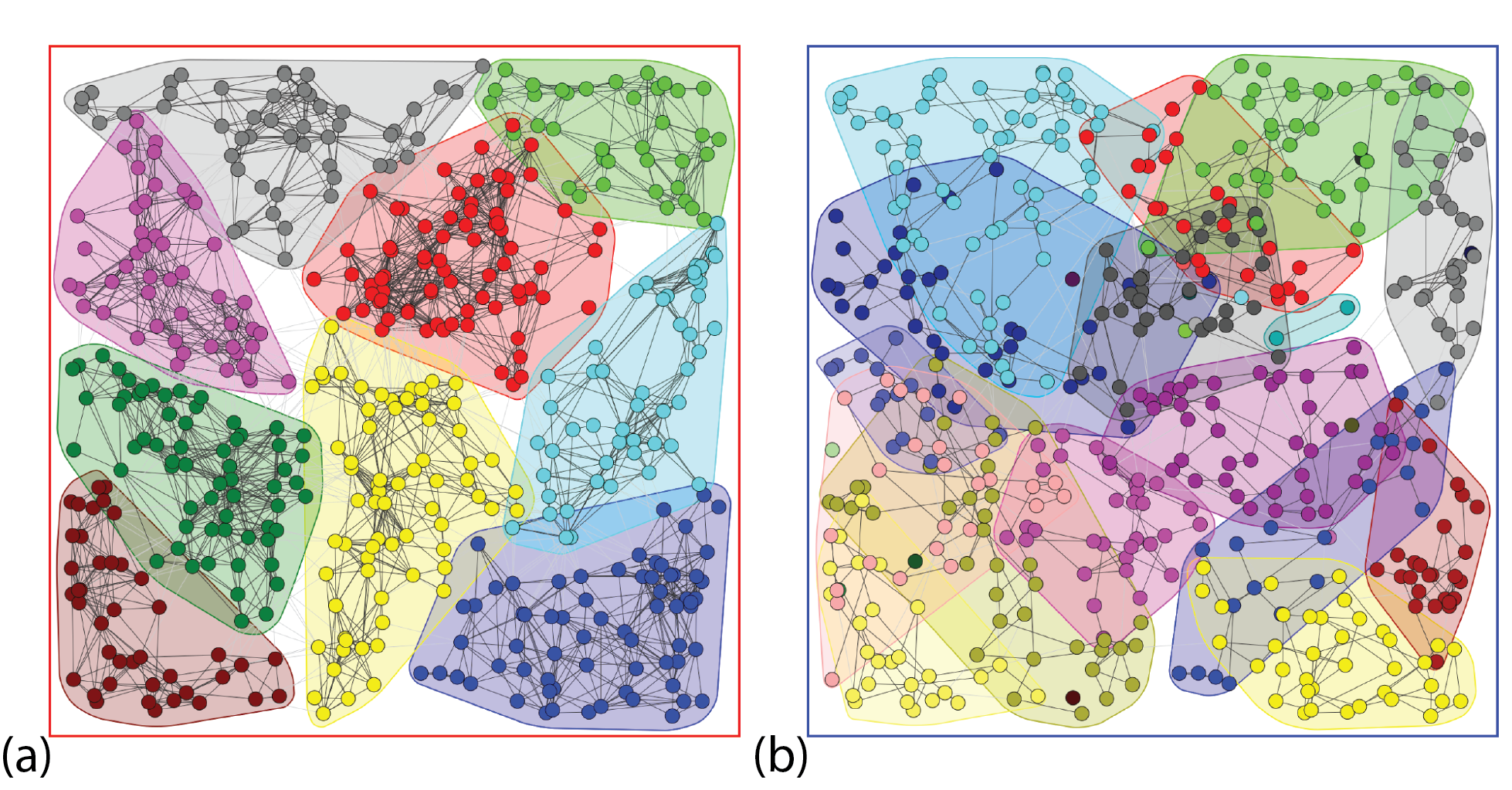}
	\caption{\textit{The partitioning of the  500 nanowire example network featured in Fig. 1 into communities based on the Louvain algorithm\cite{Blondel2008} of modularity maximization. (a) 2D and (b) Q3D.}}
	\label{fgr:comms}
\end{figure}

The separation of typical networks into modules is shown in Fig. \ref{fgr:comms}. The 2D network is much more densely connected than the Q3D network, and the communities are more clearly defined. The communities in the Q3D network can also be clearly seen, but overlap much more than in the 2D network. Fig. \ref{fgr:4panel}d shows that the modularity, $Q$ (see Methods for definition), for the Q3D network is higher for all network sizes.  For both the 2D  and Q3D networks, $Q$ is similar to that of the  regular WS networks ($p=0$) with the same dimensionality (\addlegendimageintext{black, mark=triangle*, fill=magenta, mark size=3pt, only marks, rotate=180, opacity=0.7} and \addlegendimageintext{black, mark=triangle*, fill=black!40!green, mark size=3pt, only marks, rotate=180, opacity=0.7} respectively). 

\begin{figure*}[ht]
	\centering
	\includegraphics[width=\textwidth,]{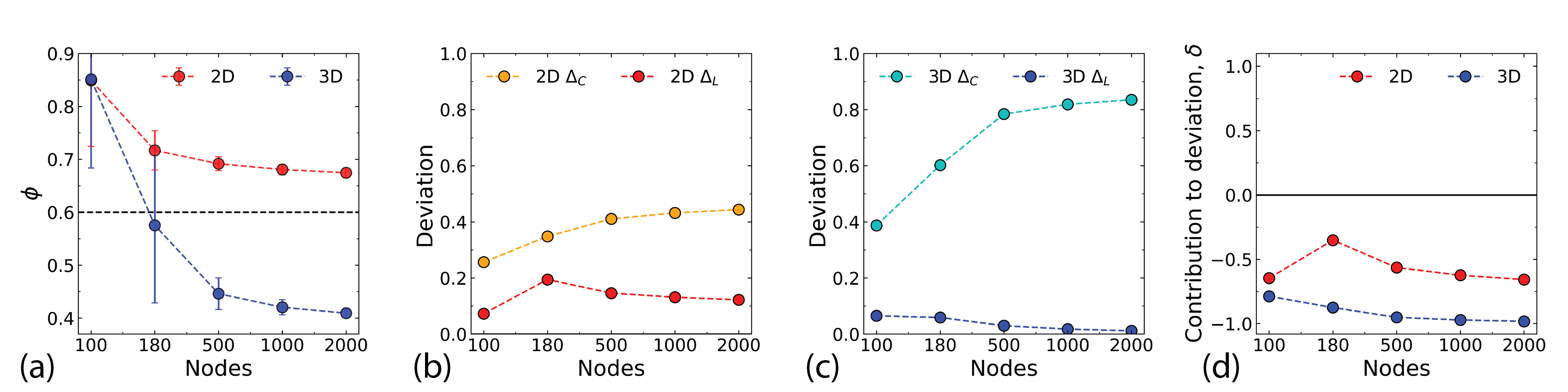}
	\caption{\textit{(a) Comparisons of the small-world propensity ($\phi$) between the 2D and Q3D networks. The averages for the networks were taken over $28$ networks, and the error bars are one standard deviation. For the larger networks, the small-world propensity of the Q3D networks is lower than that of the 2D networks. Contributions of clustering coefficient ($C$) and path length ($L$) to the small-world propensity for the 2D (b) and Q3D (c) networks. Larger values of $\Delta_C$ and $\Delta_L$ indicate that the network has large deviations from the corresponding benchmark networks (see text), leading to lower values of $\phi$. (d) $\delta$, the contribution to the deviation indicates whether $\Delta_C$ or $\Delta_L$ played a bigger role in driving the result for $\phi$. High negative values indicate that the low clustering relative to a regular network strongly drives down $\phi$ (high positive values would imply that high path lengths relative to a random network most strongly decreases the value of $\phi$). In other words, the path lengths do not deviate very much from the benchmark path lengths, but the clustering does.}}
	\label{fgr:smallworld}
\end{figure*}

The degree of modularity is of interest because it reflects the vulnerability of a network to different types of `attack' by random removal of  nodes.\cite{Shai2015} In the low modularity regime, removal of nodes fragments the modules internally and causes the network to collapse. For high modularity, the network undergoes a fragmentation process where, while the modules remain intact, they become disconnected from one another. Networks with a broader degree distribution tend to be more vulnerable to fragmentation in this way.\cite{Shai2015} This is because a broader degree distribution with the same average degree implies more low-degree nodes and hence fewer interconnections between nodes. Thus the higher modularity in the Q3D network is at least partly a consequence of the smaller average degree (see Fig. \ref{fgr:path_degree}).

\subsection*{Small-world}

A small world architecture\cite{Watts1998} is characterised by a high degree of clustering and  a small average path length between nodes, but the precise definition of the ``small-world effect" has been the subject of some debate.\cite{Newman2000} Here, we use the small-world propensity $\phi$ (see Methods for definition) to quantify the ``small-worldness" of the network.\cite{Muldoon2016}

The effects of system size on $\phi$ are shown in Fig. \ref{fgr:smallworld}a. For $N \sim 100$ the 2D and Q3D networks are almost identical because the density of the networks is low, with limited overlapping of wires. As groups of wires  begin to span the network, the small-world propensity decreases. Muldoon et al propose a pragmatic threshold value of $\phi_T = 0.6$,\cite{Muldoon2016} with networks having $\phi > \phi_T$ identified as having a strong small-world propensity. The 2D network remains consistently above this threshold ($\phi=0.67\pm0.01$ for $N=2000$). In contrast, for the Q3D networks it is clear  that for $N > N_c$  the stacking of the wires becomes important and $\phi$ drops below $\phi_T$.

Figs. \ref{fgr:smallworld}b-d further elucidate the contributions  of the clustering coefficient and the path length to $\phi$ ($\Delta_C$ and $\Delta_L$ respectively -- see methods for definitions). $\Delta_C$ is high for the Q3D network (Fig. \ref{fgr:smallworld}c) indicating a large divergence from the regular network due to lower clustering. The 2D network by comparison, has much lower values of $\Delta_C$ (Fig. \ref{fgr:smallworld}b). Both networks have low values of $\Delta_L$, due to their small path lengths, which are similar to those of the corresponding WS random networks. Fig. \ref{fgr:smallworld}d shows the contribution, $\delta$ to the deviation (see methods and ESI Fig. S7 for more details). The negative values of $\delta$ for both networks indicate that the dominant contribution to $\phi$ is the low level of clustering compared with the corresponding WS regular network. This is especially evident in the Q3D network. Ultimately, the lower small world propensity for the Q3D networks results from a more dramatic decrease in clustering than is observed for the 2D networks.

\section*{Conclusion}
We have shown that stacking of nanowires during deposition significantly affects the connectivity and topology of the resulting network. Perfectly 2D networks have much broader degree distributions and much larger mean degrees than the quasi-3D networks that result from stacking. Differences in connectivity are observable soon after the percolation threshold is exceeded i.e. as soon the number of wires is sufficient to connect opposite sides of the system (allowing current to flow through the network). More specifically the small world character of quasi-3D networks is significantly lower than in the 2D networks. The Q3D networks instead exhibit greater modularity. Perhaps surprisingly, the connectivity of the Q3D networks does not increase when additional nanowires are deposited.

These differences in connectivity will have significant effects on the properties of real-world nanowire networks, and hence on potential applications such as neuromorphic computing. It is known for example that small world connectivity is advantageous for time series prediction within a reservoir computing framework\cite{Deng2007} and significantly affects synchronisability in oscillator networks.\cite{Barahona2002, Nishikawa2003}

\section*{Methods}
The nanowire simulations were built using Python v3.7.7. The graph analysis measures were taken from the Python NetworkX\cite{networkx} and iGraph\cite{igraph} libraries, and the MATLAB Brain Connectivity Toolbox.\cite{matlab}

Since the connections between wires can be mapped to an adjacency matrix, we can use graph theory to analyze the networks. In order to simplify analysis, the corresponding graph is considered to be undirected and the edges unweighted -- that is, the adjacency matrix is a symmetric binary matrix. Each wire is modelled as a node (or vertex) in a graph, and each contact between wires is an edge. The nanowire network is then treated as the graph $G=(N,E)$, where $N$ is the set of nodes and $E$ the set of edges. The number of wires that are connected to wire $i$ is the degree, $k_i$. The mean degree $\bar{k}$ is then
\begin{equation}
\bar{k} = \frac{1}{N}\sum_i^Nk_i
\end{equation}

The degree distribution, $P(k)$, captures some aspects of the connectivity of the network, but there are many other metrics. Watts and Strogatz introduced methods to characterise the path lengths and clustering of the network, as described below.

Given any two nodes in the network $i$ and $j$, the path length $L_{ij}$, is the shortest sequence of edges that connects them. The average path length (also sometimes called the characteristic path length) is then,
\begin{equation}
	L = \frac{1}{N(N-1)}\sum_{i\ne j}^{N}L_{ij}
\end{equation}

The local clustering coefficient is the fraction of the total allowable edges that exist between node $i$ and its neighbours. For an undirected network with adjacency matrix $A$, the local clustering coefficient of node $i$ with degree $k_i$ is given by
\begin{equation}
	C_i = \frac{1}{k_i(k_i-1)}\sum_{j, k}A_{ij}A_{ik}A_{jk}
\end{equation}
The global clustering coefficient is then the average of these values
\begin{equation}
	C = \frac{1}{N}\sum_{i=1}^{N}C_i
\end{equation}

The small-world propensity, $\phi$,\cite{Muldoon2016} was defined in order to quantify the deviation of the clustering coefficient, $C_{obs}$, and path length, $L_{obs}$, of the network from those of both regular and random networks with the same number of nodes and degree distribution in such a way as to account for variations in network density:
\begin{equation}
\phi = 1 - \sqrt{ \frac{\Delta^2_C + \Delta^2_L}{2} }
\end{equation}
where,
\begin{equation}
\Delta_C = \frac{C_{latt} - C_{obs}}{C_{latt} - C_{rand}}
\end{equation}
and
\begin{equation}
\Delta_L = \frac{L_{obs} - L_{rand}}{L_{latt} - L_{rand}}
\end{equation}
In addition to the small-world propensity, following the work of Muldoon et al, we also quantify the amount that $\Delta_C$ and $\Delta_L$ contribute to $\phi$. This can be done by calculating the angular difference of the values of the network from the value of equal contribution of $\Delta_C$ and $\Delta_L$ (see ESI Fig. S7). This calculation is then called the contribution to the deviation,
\begin{equation}
\delta = \frac{4}{\pi}\tan^{-1}{\left(\frac{\Delta_L}{\Delta_C}\right)} - 1
\end{equation}

The method we used to partition the network was the Louvain method of community detection.\cite{Blondel2008} In brief, nodes within the same community are more densely connected to one another than with nodes in other communities. Each node is assigned to one community, then pairs of communities are combined iteratively such that the number of connections within groups  is maximized while the number of connections between groups is minimized. This is done by maximizing the modularity, a measure of the relative density of edges inside the groups with respect to edges outside the groups. For some partition of a network with $m$ edges with node $i$ having degree $k_i$, into communities $c$, the modularity $q$ of this is\cite{Clauset2004}
\begin{equation}
q = \frac{1}{2m}\sum_{i,j}^{n}\left( A_{ij} - \frac{k_ik_j}{2m} \right) \delta\left( c_i, c_j \right)
\end{equation}
where $\delta$ is the Kronecker delta function. The overall modularity $Q$ is then the maximum modularity of all possible partitions of the network $Q=\max (q)$. Many naturally occuring networks are found to be divisible into distinct communities via optimization of the network modularity.\cite{Newman2006}

\subsection{}
\subsubsection{}

\bibliography{nanopaper}

\begin{thebibliography}{45}%
\makeatletter
\providecommand \@ifxundefined [1]{%
 \@ifx{#1\undefined}
}%
\providecommand \@ifnum [1]{%
 \ifnum #1\expandafter \@firstoftwo
 \else \expandafter \@secondoftwo
 \fi
}%
\providecommand \@ifx [1]{%
 \ifx #1\expandafter \@firstoftwo
 \else \expandafter \@secondoftwo
 \fi
}%
\providecommand \natexlab [1]{#1}%
\providecommand \enquote  [1]{``#1''}%
\providecommand \bibnamefont  [1]{#1}%
\providecommand \bibfnamefont [1]{#1}%
\providecommand \citenamefont [1]{#1}%
\providecommand \href@noop [0]{\@secondoftwo}%
\providecommand \href [0]{\begingroup \@sanitize@url \@href}%
\providecommand \@href[1]{\@@startlink{#1}\@@href}%
\providecommand \@@href[1]{\endgroup#1\@@endlink}%
\providecommand \@sanitize@url [0]{\catcode `\\12\catcode `\$12\catcode
  `\&12\catcode `\#12\catcode `\^12\catcode `\_12\catcode `\%12\relax}%
\providecommand \@@startlink[1]{}%
\providecommand \@@endlink[0]{}%
\providecommand \url  [0]{\begingroup\@sanitize@url \@url }%
\providecommand \@url [1]{\endgroup\@href {#1}{\urlprefix }}%
\providecommand \urlprefix  [0]{URL }%
\providecommand \Eprint [0]{\href }%
\providecommand \doibase [0]{https://doi.org/}%
\providecommand \selectlanguage [0]{\@gobble}%
\providecommand \bibinfo  [0]{\@secondoftwo}%
\providecommand \bibfield  [0]{\@secondoftwo}%
\providecommand \translation [1]{[#1]}%
\providecommand \BibitemOpen [0]{}%
\providecommand \bibitemStop [0]{}%
\providecommand \bibitemNoStop [0]{.\EOS\space}%
\providecommand \EOS [0]{\spacefactor3000\relax}%
\providecommand \BibitemShut  [1]{\csname bibitem#1\endcsname}%
\let\auto@bib@innerbib\@empty
\bibitem [{\citenamefont {Ye}\ \emph {et~al.}(2014)\citenamefont {Ye},
  \citenamefont {Rathmell}, \citenamefont {Chen}, \citenamefont {Stewart},\
  and\ \citenamefont {Wiley}}]{Ye2014}%
  \BibitemOpen
  \bibfield  {author} {\bibinfo {author} {\bibfnamefont {S.}~\bibnamefont
  {Ye}}, \bibinfo {author} {\bibfnamefont {A.~R.}\ \bibnamefont {Rathmell}},
  \bibinfo {author} {\bibfnamefont {Z.}~\bibnamefont {Chen}}, \bibinfo {author}
  {\bibfnamefont {I.~E.}\ \bibnamefont {Stewart}},\ and\ \bibinfo {author}
  {\bibfnamefont {B.~J.}\ \bibnamefont {Wiley}},\ }\bibfield  {title} {\bibinfo
  {title} {Metal nanowire networks: The next generation of transparent
  conductors},\ }\href {https://doi.org/10.1002/adma.201402710} {\bibfield
  {journal} {\bibinfo  {journal} {Advanced Materials}\ }\textbf {\bibinfo
  {volume} {26}},\ \bibinfo {pages} {6670} (\bibinfo {year}
  {2014})}\BibitemShut {NoStop}%
\bibitem [{\citenamefont {Park}\ \emph {et~al.}(2013)\citenamefont {Park},
  \citenamefont {Vosguerichian},\ and\ \citenamefont {Bao}}]{Park2013}%
  \BibitemOpen
  \bibfield  {author} {\bibinfo {author} {\bibfnamefont {S.}~\bibnamefont
  {Park}}, \bibinfo {author} {\bibfnamefont {M.}~\bibnamefont
  {Vosguerichian}},\ and\ \bibinfo {author} {\bibfnamefont {Z.}~\bibnamefont
  {Bao}},\ }\bibfield  {title} {\bibinfo {title} {A review of fabrication and
  applications of carbon nanotube film-based flexible electronics},\ }\href
  {https://doi.org/10.1039/C3NR33560G} {\bibfield  {journal} {\bibinfo
  {journal} {Nanoscale}\ }\textbf {\bibinfo {volume} {5}},\ \bibinfo {pages}
  {1727} (\bibinfo {year} {2013})}\BibitemShut {NoStop}%
\bibitem [{\citenamefont {Sannicolo}\ \emph {et~al.}(2016)\citenamefont
  {Sannicolo}, \citenamefont {Lagrange}, \citenamefont {Cabos}, \citenamefont
  {Celle}, \citenamefont {Simonato},\ and\ \citenamefont
  {Bellet}}]{Sannicolo2016}%
  \BibitemOpen
  \bibfield  {author} {\bibinfo {author} {\bibfnamefont {T.}~\bibnamefont
  {Sannicolo}}, \bibinfo {author} {\bibfnamefont {M.}~\bibnamefont {Lagrange}},
  \bibinfo {author} {\bibfnamefont {A.}~\bibnamefont {Cabos}}, \bibinfo
  {author} {\bibfnamefont {C.}~\bibnamefont {Celle}}, \bibinfo {author}
  {\bibfnamefont {J.-P.}\ \bibnamefont {Simonato}},\ and\ \bibinfo {author}
  {\bibfnamefont {D.}~\bibnamefont {Bellet}},\ }\bibfield  {title} {\bibinfo
  {title} {Metallic nanowire-based transparent electrodes for next generation
  flexible devices: a review},\ }\href {https://doi.org/10.1002/smll.201602581}
  {\bibfield  {journal} {\bibinfo  {journal} {Small}\ }\textbf {\bibinfo
  {volume} {12}},\ \bibinfo {pages} {6052} (\bibinfo {year}
  {2016})}\BibitemShut {NoStop}%
\bibitem [{\citenamefont {Tian}\ and\ \citenamefont {Lieber}(2019)}]{Tian2019}%
  \BibitemOpen
  \bibfield  {author} {\bibinfo {author} {\bibfnamefont {B.}~\bibnamefont
  {Tian}}\ and\ \bibinfo {author} {\bibfnamefont {C.~M.}\ \bibnamefont
  {Lieber}},\ }\bibfield  {title} {\bibinfo {title} {Nanowired bioelectric
  interfaces},\ }\href {https://doi.org/10.1021/acs.chemrev.8b00795} {\bibfield
   {journal} {\bibinfo  {journal} {Chemical Reviews}\ }\textbf {\bibinfo
  {volume} {119}},\ \bibinfo {pages} {9136} (\bibinfo {year} {2019})},\
  \bibinfo {note} {pMID: 30995019}\BibitemShut {NoStop}%
\bibitem [{\citenamefont {Fennell~Jr.}\ \emph {et~al.}(2016)\citenamefont
  {Fennell~Jr.}, \citenamefont {Liu}, \citenamefont {Azzarelli}, \citenamefont
  {Weis}, \citenamefont {Rochat}, \citenamefont {Mirica}, \citenamefont
  {Ravnsbæk},\ and\ \citenamefont {Swager}}]{Fennell2016}%
  \BibitemOpen
  \bibfield  {author} {\bibinfo {author} {\bibfnamefont {J.~F.}\ \bibnamefont
  {Fennell~Jr.}}, \bibinfo {author} {\bibfnamefont {S.~F.}\ \bibnamefont
  {Liu}}, \bibinfo {author} {\bibfnamefont {J.~M.}\ \bibnamefont {Azzarelli}},
  \bibinfo {author} {\bibfnamefont {J.~G.}\ \bibnamefont {Weis}}, \bibinfo
  {author} {\bibfnamefont {S.}~\bibnamefont {Rochat}}, \bibinfo {author}
  {\bibfnamefont {K.~A.}\ \bibnamefont {Mirica}}, \bibinfo {author}
  {\bibfnamefont {J.~B.}\ \bibnamefont {Ravnsbæk}},\ and\ \bibinfo {author}
  {\bibfnamefont {T.~M.}\ \bibnamefont {Swager}},\ }\bibfield  {title}
  {\bibinfo {title} {Nanowire chemical/biological sensors: Status and a roadmap
  for the future},\ }\href {https://doi.org/10.1002/anie.201505308} {\bibfield
  {journal} {\bibinfo  {journal} {Angewandte Chemie International Edition}\
  }\textbf {\bibinfo {volume} {55}},\ \bibinfo {pages} {1266} (\bibinfo {year}
  {2016})}\BibitemShut {NoStop}%
\bibitem [{\citenamefont {Zheng}\ \emph {et~al.}(2016)\citenamefont {Zheng},
  \citenamefont {Alsager}, \citenamefont {Zhu}, \citenamefont {Travas-Sejdic},
  \citenamefont {Hodgkiss},\ and\ \citenamefont {Plank}}]{Zheng2016}%
  \BibitemOpen
  \bibfield  {author} {\bibinfo {author} {\bibfnamefont {H.~Y.}\ \bibnamefont
  {Zheng}}, \bibinfo {author} {\bibfnamefont {O.~A.}\ \bibnamefont {Alsager}},
  \bibinfo {author} {\bibfnamefont {B.}~\bibnamefont {Zhu}}, \bibinfo {author}
  {\bibfnamefont {J.}~\bibnamefont {Travas-Sejdic}}, \bibinfo {author}
  {\bibfnamefont {J.~M.}\ \bibnamefont {Hodgkiss}},\ and\ \bibinfo {author}
  {\bibfnamefont {N.~O.~V.}\ \bibnamefont {Plank}},\ }\bibfield  {title}
  {\bibinfo {title} {Electrostatic gating in carbon nanotube aptasensors},\
  }\href {https://doi.org/10.1039/C5NR08117C} {\bibfield  {journal} {\bibinfo
  {journal} {Nanoscale}\ }\textbf {\bibinfo {volume} {8}},\ \bibinfo {pages}
  {13659} (\bibinfo {year} {2016})}\BibitemShut {NoStop}%
\bibitem [{\citenamefont {Song}\ \emph {et~al.}(2016)\citenamefont {Song},
  \citenamefont {Han}, \citenamefont {Ju},\ and\ \citenamefont
  {Kim}}]{Song2016}%
  \BibitemOpen
  \bibfield  {author} {\bibinfo {author} {\bibfnamefont {C.-H.}\ \bibnamefont
  {Song}}, \bibinfo {author} {\bibfnamefont {C.~J.}\ \bibnamefont {Han}},
  \bibinfo {author} {\bibfnamefont {B.-K.}\ \bibnamefont {Ju}},\ and\ \bibinfo
  {author} {\bibfnamefont {J.-W.}\ \bibnamefont {Kim}},\ }\bibfield  {title}
  {\bibinfo {title} {Photoenhanced patterning of metal nanowire networks for
  fabrication of ultraflexible transparent devices},\ }\href
  {https://doi.org/10.1021/acsami.5b09386} {\bibfield  {journal} {\bibinfo
  {journal} {ACS Applied Materials and Interfaces}\ }\textbf {\bibinfo {volume}
  {8}},\ \bibinfo {pages} {480} (\bibinfo {year} {2016})}\BibitemShut {NoStop}%
\bibitem [{\citenamefont {Lee}\ \emph {et~al.}(2017)\citenamefont {Lee},
  \citenamefont {An}, \citenamefont {Won}, \citenamefont {Ka}, \citenamefont
  {Hwang}, \citenamefont {Moon}, \citenamefont {Kwon}, \citenamefont {Hong},
  \citenamefont {Kim}, \citenamefont {Lee},\ and\ \citenamefont
  {Ko}}]{Lee2017}%
  \BibitemOpen
  \bibfield  {author} {\bibinfo {author} {\bibfnamefont {J.}~\bibnamefont
  {Lee}}, \bibinfo {author} {\bibfnamefont {K.}~\bibnamefont {An}}, \bibinfo
  {author} {\bibfnamefont {P.}~\bibnamefont {Won}}, \bibinfo {author}
  {\bibfnamefont {Y.}~\bibnamefont {Ka}}, \bibinfo {author} {\bibfnamefont
  {H.}~\bibnamefont {Hwang}}, \bibinfo {author} {\bibfnamefont
  {H.}~\bibnamefont {Moon}}, \bibinfo {author} {\bibfnamefont {Y.}~\bibnamefont
  {Kwon}}, \bibinfo {author} {\bibfnamefont {S.}~\bibnamefont {Hong}}, \bibinfo
  {author} {\bibfnamefont {C.}~\bibnamefont {Kim}}, \bibinfo {author}
  {\bibfnamefont {C.}~\bibnamefont {Lee}},\ and\ \bibinfo {author}
  {\bibfnamefont {S.~H.}\ \bibnamefont {Ko}},\ }\bibfield  {title} {\bibinfo
  {title} {A dual-scale metal nanowire network transparent conductor for highly
  efficient and flexible organic light emitting diodes},\ }\href
  {https://doi.org/10.1039/C6NR09902E} {\bibfield  {journal} {\bibinfo
  {journal} {Nanoscale}\ }\textbf {\bibinfo {volume} {9}},\ \bibinfo {pages}
  {1978} (\bibinfo {year} {2017})}\BibitemShut {NoStop}%
\bibitem [{\citenamefont {Li}\ \emph {et~al.}(2020)\citenamefont {Li},
  \citenamefont {Zhang}, \citenamefont {Shi}, \citenamefont {Xu}, \citenamefont
  {Qin}, \citenamefont {He}, \citenamefont {Yang}, \citenamefont {Dai},
  \citenamefont {Liu}, \citenamefont {Zhou}, \citenamefont {Yu}, \citenamefont
  {Silva},\ and\ \citenamefont {Fahlman}}]{Li2020}%
  \BibitemOpen
  \bibfield  {author} {\bibinfo {author} {\bibfnamefont {W.}~\bibnamefont
  {Li}}, \bibinfo {author} {\bibfnamefont {H.}~\bibnamefont {Zhang}}, \bibinfo
  {author} {\bibfnamefont {S.}~\bibnamefont {Shi}}, \bibinfo {author}
  {\bibfnamefont {J.}~\bibnamefont {Xu}}, \bibinfo {author} {\bibfnamefont
  {X.}~\bibnamefont {Qin}}, \bibinfo {author} {\bibfnamefont {Q.}~\bibnamefont
  {He}}, \bibinfo {author} {\bibfnamefont {K.}~\bibnamefont {Yang}}, \bibinfo
  {author} {\bibfnamefont {W.}~\bibnamefont {Dai}}, \bibinfo {author}
  {\bibfnamefont {G.}~\bibnamefont {Liu}}, \bibinfo {author} {\bibfnamefont
  {Q.}~\bibnamefont {Zhou}}, \bibinfo {author} {\bibfnamefont {H.}~\bibnamefont
  {Yu}}, \bibinfo {author} {\bibfnamefont {S.~R.~P.}\ \bibnamefont {Silva}},\
  and\ \bibinfo {author} {\bibfnamefont {M.}~\bibnamefont {Fahlman}},\
  }\bibfield  {title} {\bibinfo {title} {Recent progress in silver nanowire
  networks for flexible organic electronics},\ }\href
  {https://doi.org/10.1039/C9TC06865A} {\bibfield  {journal} {\bibinfo
  {journal} {J. Mater. Chem. C}\ }\textbf {\bibinfo {volume} {8}},\ \bibinfo
  {pages} {4636} (\bibinfo {year} {2020})}\BibitemShut {NoStop}%
\bibitem [{\citenamefont {Mallinson}\ \emph {et~al.}(2019)\citenamefont
  {Mallinson}, \citenamefont {Shirai}, \citenamefont {Acharya}, \citenamefont
  {Bose}, \citenamefont {Galli},\ and\ \citenamefont {Brown}}]{Mallinson2019}%
  \BibitemOpen
  \bibfield  {author} {\bibinfo {author} {\bibfnamefont {J.~B.}\ \bibnamefont
  {Mallinson}}, \bibinfo {author} {\bibfnamefont {S.}~\bibnamefont {Shirai}},
  \bibinfo {author} {\bibfnamefont {S.~K.}\ \bibnamefont {Acharya}}, \bibinfo
  {author} {\bibfnamefont {S.~K.}\ \bibnamefont {Bose}}, \bibinfo {author}
  {\bibfnamefont {E.}~\bibnamefont {Galli}},\ and\ \bibinfo {author}
  {\bibfnamefont {S.~A.}\ \bibnamefont {Brown}},\ }\bibfield  {title} {\bibinfo
  {title} {Avalanches and criticality in self-organized nanoscale networks},\
  }\href {https://doi.org/10.1126/sciadv.aaw8438} {\bibfield  {journal}
  {\bibinfo  {journal} {Science Advances}\ }\textbf {\bibinfo {volume} {5}},\
  \bibinfo {pages} {10.1126/sciadv.aaw8438} (\bibinfo {year}
  {2019})}\BibitemShut {NoStop}%
\bibitem [{\citenamefont {Shirai}\ \emph {et~al.}(2020)\citenamefont {Shirai},
  \citenamefont {Acharya}, \citenamefont {Bose}, \citenamefont {Mallinson},
  \citenamefont {Galli}, \citenamefont {Pike}, \citenamefont {Arnold},\ and\
  \citenamefont {Brown}}]{Shirai2020}%
  \BibitemOpen
  \bibfield  {author} {\bibinfo {author} {\bibfnamefont {S.}~\bibnamefont
  {Shirai}}, \bibinfo {author} {\bibfnamefont {S.~K.}\ \bibnamefont {Acharya}},
  \bibinfo {author} {\bibfnamefont {S.~K.}\ \bibnamefont {Bose}}, \bibinfo
  {author} {\bibfnamefont {J.~B.}\ \bibnamefont {Mallinson}}, \bibinfo {author}
  {\bibfnamefont {E.}~\bibnamefont {Galli}}, \bibinfo {author} {\bibfnamefont
  {M.~D.}\ \bibnamefont {Pike}}, \bibinfo {author} {\bibfnamefont {M.~D.}\
  \bibnamefont {Arnold}},\ and\ \bibinfo {author} {\bibfnamefont {S.~A.}\
  \bibnamefont {Brown}},\ }\bibfield  {title} {\bibinfo {title} {Long-range
  temporal correlations in scale-free neuromorphic networks},\ }\href
  {https://doi.org/10.1162/netn\_a\_00128} {\bibfield  {journal} {\bibinfo
  {journal} {Network Neuroscience}\ }\textbf {\bibinfo {volume} {4}},\ \bibinfo
  {pages} {432} (\bibinfo {year} {2020})}\BibitemShut {NoStop}%
\bibitem [{\citenamefont {Pike}\ \emph {et~al.}(2020)\citenamefont {Pike},
  \citenamefont {Bose}, \citenamefont {Mallinson}, \citenamefont {Acharya},
  \citenamefont {Shirai}, \citenamefont {Galli}, \citenamefont {Weddell},
  \citenamefont {Bones}, \citenamefont {Arnold},\ and\ \citenamefont
  {Brown}}]{Pike2020}%
  \BibitemOpen
  \bibfield  {author} {\bibinfo {author} {\bibfnamefont {M.~D.}\ \bibnamefont
  {Pike}}, \bibinfo {author} {\bibfnamefont {S.~K.}\ \bibnamefont {Bose}},
  \bibinfo {author} {\bibfnamefont {J.~B.}\ \bibnamefont {Mallinson}}, \bibinfo
  {author} {\bibfnamefont {S.~K.}\ \bibnamefont {Acharya}}, \bibinfo {author}
  {\bibfnamefont {S.}~\bibnamefont {Shirai}}, \bibinfo {author} {\bibfnamefont
  {E.}~\bibnamefont {Galli}}, \bibinfo {author} {\bibfnamefont {S.~J.}\
  \bibnamefont {Weddell}}, \bibinfo {author} {\bibfnamefont {P.~J.}\
  \bibnamefont {Bones}}, \bibinfo {author} {\bibfnamefont {M.~D.}\ \bibnamefont
  {Arnold}},\ and\ \bibinfo {author} {\bibfnamefont {S.~A.}\ \bibnamefont
  {Brown}},\ }\bibfield  {title} {\bibinfo {title} {{Atomic Scale Dynamics
  Drive Brain-like Avalanches in Percolating Nanostructured Networks}},\ }\href
  {https://doi.org/10.1021/acs.nanolett.0c01096} {\bibfield  {journal}
  {\bibinfo  {journal} {Nano Letters}\ }\textbf {\bibinfo {volume} {20}},\
  \bibinfo {pages} {3935} (\bibinfo {year} {2020})}\BibitemShut {NoStop}%
\bibitem [{\citenamefont {Avizienis}\ \emph {et~al.}(2012)\citenamefont
  {Avizienis}, \citenamefont {Sillin}, \citenamefont {Martin-Olmos},
  \citenamefont {Shieh}, \citenamefont {Aono}, \citenamefont {Stieg},\ and\
  \citenamefont {Gimzewski}}]{Avizienis2012}%
  \BibitemOpen
  \bibfield  {author} {\bibinfo {author} {\bibfnamefont {A.~V.}\ \bibnamefont
  {Avizienis}}, \bibinfo {author} {\bibfnamefont {H.~O.}\ \bibnamefont
  {Sillin}}, \bibinfo {author} {\bibfnamefont {C.}~\bibnamefont
  {Martin-Olmos}}, \bibinfo {author} {\bibfnamefont {H.~H.}\ \bibnamefont
  {Shieh}}, \bibinfo {author} {\bibfnamefont {M.}~\bibnamefont {Aono}},
  \bibinfo {author} {\bibfnamefont {A.~Z.}\ \bibnamefont {Stieg}},\ and\
  \bibinfo {author} {\bibfnamefont {J.~K.}\ \bibnamefont {Gimzewski}},\
  }\bibfield  {title} {\bibinfo {title} {Neuromorphic atomic switch networks},\
  }\href {https://doi.org/10.1371/journal.pone.0042772} {\bibfield  {journal}
  {\bibinfo  {journal} {PLOS ONE}\ }\textbf {\bibinfo {volume} {7}},\ \bibinfo
  {pages} {1} (\bibinfo {year} {2012})}\BibitemShut {NoStop}%
\bibitem [{\citenamefont {Manning}\ \emph {et~al.}(2018)\citenamefont
  {Manning}, \citenamefont {Niosi}, \citenamefont {da~Rocha}, \citenamefont
  {Bellew}, \citenamefont {O'Callaghan}, \citenamefont {Biswas}, \citenamefont
  {Flowers}, \citenamefont {Wiley}, \citenamefont {Holmes}, \citenamefont
  {Ferreira},\ and\ \citenamefont {Boland}}]{Manning2018}%
  \BibitemOpen
  \bibfield  {author} {\bibinfo {author} {\bibfnamefont {H.~G.}\ \bibnamefont
  {Manning}}, \bibinfo {author} {\bibfnamefont {F.}~\bibnamefont {Niosi}},
  \bibinfo {author} {\bibfnamefont {C.~G.}\ \bibnamefont {da~Rocha}}, \bibinfo
  {author} {\bibfnamefont {A.~T.}\ \bibnamefont {Bellew}}, \bibinfo {author}
  {\bibfnamefont {C.}~\bibnamefont {O'Callaghan}}, \bibinfo {author}
  {\bibfnamefont {S.}~\bibnamefont {Biswas}}, \bibinfo {author} {\bibfnamefont
  {P.~F.}\ \bibnamefont {Flowers}}, \bibinfo {author} {\bibfnamefont {B.~J.}\
  \bibnamefont {Wiley}}, \bibinfo {author} {\bibfnamefont {J.~D.}\ \bibnamefont
  {Holmes}}, \bibinfo {author} {\bibfnamefont {M.~S.}\ \bibnamefont
  {Ferreira}},\ and\ \bibinfo {author} {\bibfnamefont {J.~J.}\ \bibnamefont
  {Boland}},\ }\bibfield  {title} {\bibinfo {title} {{Emergence of
  winner-takes-all connectivity paths in random nanowire networks}},\ }\href
  {https://doi.org/10.1038/s41467-018-05517-6} {\bibfield  {journal} {\bibinfo
  {journal} {Nature Communications}\ }\textbf {\bibinfo {volume} {9}},\
  \bibinfo {pages} {3219} (\bibinfo {year} {2018})}\BibitemShut {NoStop}%
\bibitem [{\citenamefont {Tanaka}\ \emph {et~al.}(2018)\citenamefont {Tanaka},
  \citenamefont {Akai-Kasaya}, \citenamefont {TermehYousefi}, \citenamefont
  {Hong}, \citenamefont {Fu}, \citenamefont {Tamukoh}, \citenamefont {Tanaka},
  \citenamefont {Asai},\ and\ \citenamefont {Ogawa}}]{Tanaka2018}%
  \BibitemOpen
  \bibfield  {author} {\bibinfo {author} {\bibfnamefont {H.}~\bibnamefont
  {Tanaka}}, \bibinfo {author} {\bibfnamefont {M.}~\bibnamefont {Akai-Kasaya}},
  \bibinfo {author} {\bibfnamefont {A.}~\bibnamefont {TermehYousefi}}, \bibinfo
  {author} {\bibfnamefont {L.}~\bibnamefont {Hong}}, \bibinfo {author}
  {\bibfnamefont {L.}~\bibnamefont {Fu}}, \bibinfo {author} {\bibfnamefont
  {H.}~\bibnamefont {Tamukoh}}, \bibinfo {author} {\bibfnamefont
  {D.}~\bibnamefont {Tanaka}}, \bibinfo {author} {\bibfnamefont
  {T.}~\bibnamefont {Asai}},\ and\ \bibinfo {author} {\bibfnamefont
  {T.}~\bibnamefont {Ogawa}},\ }\bibfield  {title} {\bibinfo {title} {{A
  molecular neuromorphic network device consisting of single-walled carbon
  nanotubes complexed with polyoxometalate}},\ }\href
  {https://doi.org/10.1038/s41467-018-04886-2} {\bibfield  {journal} {\bibinfo
  {journal} {Nature Communications}\ }\textbf {\bibinfo {volume} {9}},\
  \bibinfo {pages} {2693} (\bibinfo {year} {2018})}\BibitemShut {NoStop}%
\bibitem [{\citenamefont {Stieg}\ \emph {et~al.}(2012)\citenamefont {Stieg},
  \citenamefont {Avizienis}, \citenamefont {Sillin}, \citenamefont
  {Martin-Olmos}, \citenamefont {Aono},\ and\ \citenamefont
  {Gimzewski}}]{Stieg2012}%
  \BibitemOpen
  \bibfield  {author} {\bibinfo {author} {\bibfnamefont {A.~Z.}\ \bibnamefont
  {Stieg}}, \bibinfo {author} {\bibfnamefont {A.~V.}\ \bibnamefont
  {Avizienis}}, \bibinfo {author} {\bibfnamefont {H.~O.}\ \bibnamefont
  {Sillin}}, \bibinfo {author} {\bibfnamefont {C.}~\bibnamefont
  {Martin-Olmos}}, \bibinfo {author} {\bibfnamefont {M.}~\bibnamefont {Aono}},\
  and\ \bibinfo {author} {\bibfnamefont {J.~K.}\ \bibnamefont {Gimzewski}},\
  }\bibfield  {title} {\bibinfo {title} {Emergent criticality in complex turing
  b-type atomic switch networks},\ }\href
  {https://doi.org/10.1002/adma.201103053} {\bibfield  {journal} {\bibinfo
  {journal} {Advanced Materials}\ }\textbf {\bibinfo {volume} {24}},\ \bibinfo
  {pages} {286} (\bibinfo {year} {2012})}\BibitemShut {NoStop}%
\bibitem [{\citenamefont {Diaz-Alvarez}\ \emph {et~al.}(2019)\citenamefont
  {Diaz-Alvarez}, \citenamefont {Higuchi}, \citenamefont {Sanz-Leon},
  \citenamefont {Marcus}, \citenamefont {Shingaya}, \citenamefont {Stieg},
  \citenamefont {Gimzewski}, \citenamefont {Kuncic},\ and\ \citenamefont
  {Nakayama}}]{Diaz2019}%
  \BibitemOpen
  \bibfield  {author} {\bibinfo {author} {\bibfnamefont {A.}~\bibnamefont
  {Diaz-Alvarez}}, \bibinfo {author} {\bibfnamefont {R.}~\bibnamefont
  {Higuchi}}, \bibinfo {author} {\bibfnamefont {P.}~\bibnamefont {Sanz-Leon}},
  \bibinfo {author} {\bibfnamefont {I.}~\bibnamefont {Marcus}}, \bibinfo
  {author} {\bibfnamefont {Y.}~\bibnamefont {Shingaya}}, \bibinfo {author}
  {\bibfnamefont {A.~Z.}\ \bibnamefont {Stieg}}, \bibinfo {author}
  {\bibfnamefont {J.~K.}\ \bibnamefont {Gimzewski}}, \bibinfo {author}
  {\bibfnamefont {Z.}~\bibnamefont {Kuncic}},\ and\ \bibinfo {author}
  {\bibfnamefont {T.}~\bibnamefont {Nakayama}},\ }\bibfield  {title} {\bibinfo
  {title} {Emergent dynamics of neuromorphic nanowire networks},\ }\href
  {https://doi.org/10.1038/s41598-019-51330-6} {\bibfield  {journal} {\bibinfo
  {journal} {Scientific Reports}\ }\textbf {\bibinfo {volume} {9}},\ \bibinfo
  {pages} {14920} (\bibinfo {year} {2019})}\BibitemShut {NoStop}%
\bibitem [{\citenamefont {Milano}\ \emph {et~al.}(2020)\citenamefont {Milano},
  \citenamefont {Pedretti}, \citenamefont {Fretto}, \citenamefont {Boarino},
  \citenamefont {Benfenati}, \citenamefont {Ielmini}, \citenamefont {Valov},\
  and\ \citenamefont {Ricciardi}}]{Milano2020}%
  \BibitemOpen
  \bibfield  {author} {\bibinfo {author} {\bibfnamefont {G.}~\bibnamefont
  {Milano}}, \bibinfo {author} {\bibfnamefont {G.}~\bibnamefont {Pedretti}},
  \bibinfo {author} {\bibfnamefont {M.}~\bibnamefont {Fretto}}, \bibinfo
  {author} {\bibfnamefont {L.}~\bibnamefont {Boarino}}, \bibinfo {author}
  {\bibfnamefont {F.}~\bibnamefont {Benfenati}}, \bibinfo {author}
  {\bibfnamefont {D.}~\bibnamefont {Ielmini}}, \bibinfo {author} {\bibfnamefont
  {I.}~\bibnamefont {Valov}},\ and\ \bibinfo {author} {\bibfnamefont
  {C.}~\bibnamefont {Ricciardi}},\ }\bibfield  {title} {\bibinfo {title}
  {Brain-inspired structural plasticity through reweighting and rewiring in
  multi-terminal self-organizing memristive nanowire networks},\ }\href
  {https://doi.org/10.1002/aisy.202000096} {\bibfield  {journal} {\bibinfo
  {journal} {Advanced Intelligent Systems}\ }\textbf {\bibinfo {volume} {2}},\
  \bibinfo {pages} {2000096} (\bibinfo {year} {2020})}\BibitemShut {NoStop}%
\bibitem [{\citenamefont {Fu}\ \emph {et~al.}(2020)\citenamefont {Fu},
  \citenamefont {Zhu}, \citenamefont {Loeffler}, \citenamefont {Hochstetter},
  \citenamefont {Diaz-Alvarez}, \citenamefont {Stieg}, \citenamefont
  {Gimzewski}, \citenamefont {Nakayama},\ and\ \citenamefont
  {Kuncic}}]{Fu2020}%
  \BibitemOpen
  \bibfield  {author} {\bibinfo {author} {\bibfnamefont {K.}~\bibnamefont
  {Fu}}, \bibinfo {author} {\bibfnamefont {R.}~\bibnamefont {Zhu}}, \bibinfo
  {author} {\bibfnamefont {A.}~\bibnamefont {Loeffler}}, \bibinfo {author}
  {\bibfnamefont {J.}~\bibnamefont {Hochstetter}}, \bibinfo {author}
  {\bibfnamefont {A.}~\bibnamefont {Diaz-Alvarez}}, \bibinfo {author}
  {\bibfnamefont {A.}~\bibnamefont {Stieg}}, \bibinfo {author} {\bibfnamefont
  {J.}~\bibnamefont {Gimzewski}}, \bibinfo {author} {\bibfnamefont
  {T.}~\bibnamefont {Nakayama}},\ and\ \bibinfo {author} {\bibfnamefont
  {Z.}~\bibnamefont {Kuncic}},\ }\bibfield  {title} {\bibinfo {title}
  {Reservoir computing with neuromemristive nanowire networks},\ }in\ \href
  {https://doi.org/10.1109/IJCNN48605.2020.9207727} {\emph {\bibinfo
  {booktitle} {2020 International Joint Conference on Neural Networks
  (IJCNN)}}}\ (\bibinfo {year} {2020})\ pp.\ \bibinfo {pages}
  {1--8}\BibitemShut {NoStop}%
\bibitem [{\citenamefont {{Kuncic}}\ \emph {et~al.}(2020)\citenamefont
  {{Kuncic}}, \citenamefont {{Kavehei}}, \citenamefont {{Zhu}}, \citenamefont
  {{Loeffler}}, \citenamefont {{Fu}}, \citenamefont {{Hochstetter}},
  \citenamefont {{Li}}, \citenamefont {{Shine}}, \citenamefont
  {{Diaz-Alvarez}}, \citenamefont {{Stieg}}, \citenamefont {{Gimzewski}},\ and\
  \citenamefont {{Nakayama}}}]{Kuncic2020}%
  \BibitemOpen
  \bibfield  {author} {\bibinfo {author} {\bibfnamefont {Z.}~\bibnamefont
  {{Kuncic}}}, \bibinfo {author} {\bibfnamefont {O.}~\bibnamefont {{Kavehei}}},
  \bibinfo {author} {\bibfnamefont {R.}~\bibnamefont {{Zhu}}}, \bibinfo
  {author} {\bibfnamefont {A.}~\bibnamefont {{Loeffler}}}, \bibinfo {author}
  {\bibfnamefont {K.}~\bibnamefont {{Fu}}}, \bibinfo {author} {\bibfnamefont
  {J.}~\bibnamefont {{Hochstetter}}}, \bibinfo {author} {\bibfnamefont
  {M.}~\bibnamefont {{Li}}}, \bibinfo {author} {\bibfnamefont {J.~M.}\
  \bibnamefont {{Shine}}}, \bibinfo {author} {\bibfnamefont {A.}~\bibnamefont
  {{Diaz-Alvarez}}}, \bibinfo {author} {\bibfnamefont {A.}~\bibnamefont
  {{Stieg}}}, \bibinfo {author} {\bibfnamefont {J.}~\bibnamefont
  {{Gimzewski}}},\ and\ \bibinfo {author} {\bibfnamefont {T.}~\bibnamefont
  {{Nakayama}}},\ }\bibfield  {title} {\bibinfo {title} {Neuromorphic
  information processing with nanowire networks},\ }in\ \href
  {https://doi.org/10.1109/ISCAS45731.2020.9181034} {\emph {\bibinfo
  {booktitle} {2020 IEEE International Symposium on Circuits and Systems
  (ISCAS)}}}\ (\bibinfo {year} {2020})\ pp.\ \bibinfo {pages}
  {1--5}\BibitemShut {NoStop}%
\bibitem [{\citenamefont {Stauffer}\ and\ \citenamefont
  {Aharony}(1992)}]{Stauffer1992}%
  \BibitemOpen
  \bibfield  {author} {\bibinfo {author} {\bibfnamefont {D.}~\bibnamefont
  {Stauffer}}\ and\ \bibinfo {author} {\bibfnamefont {A.}~\bibnamefont
  {Aharony}},\ }\href {https://doi.org/10.1201/9781315274386} {\emph {\bibinfo
  {title} {Introduction to Percolation Theory}}},\ \bibinfo {edition} {2nd}\
  ed.\ (\bibinfo  {publisher} {Taylor and Francis},\ \bibinfo {year}
  {1992})\BibitemShut {NoStop}%
\bibitem [{\citenamefont {Fostner}\ \emph {et~al.}(2014)\citenamefont
  {Fostner}, \citenamefont {Brown}, \citenamefont {Carr},\ and\ \citenamefont
  {Brown}}]{Fostner2014}%
  \BibitemOpen
  \bibfield  {author} {\bibinfo {author} {\bibfnamefont {S.}~\bibnamefont
  {Fostner}}, \bibinfo {author} {\bibfnamefont {R.}~\bibnamefont {Brown}},
  \bibinfo {author} {\bibfnamefont {J.}~\bibnamefont {Carr}},\ and\ \bibinfo
  {author} {\bibfnamefont {S.~A.}\ \bibnamefont {Brown}},\ }\bibfield  {title}
  {\bibinfo {title} {Continuum percolation with tunneling},\ }\href
  {https://doi.org/10.1103/PhysRevB.89.075402} {\bibfield  {journal} {\bibinfo
  {journal} {Phys. Rev. B}\ }\textbf {\bibinfo {volume} {89}},\ \bibinfo
  {pages} {075402} (\bibinfo {year} {2014})}\BibitemShut {NoStop}%
\bibitem [{\citenamefont {Li}\ and\ \citenamefont {Zhang}(2009)}]{Li2009}%
  \BibitemOpen
  \bibfield  {author} {\bibinfo {author} {\bibfnamefont {J.}~\bibnamefont
  {Li}}\ and\ \bibinfo {author} {\bibfnamefont {S.-L.}\ \bibnamefont {Zhang}},\
  }\bibfield  {title} {\bibinfo {title} {Finite-size scaling in stick
  percolation},\ }\href@noop {} {\bibfield  {journal} {\bibinfo  {journal}
  {Physical Review E}\ }\textbf {\bibinfo {volume} {80}},\ \bibinfo {pages}
  {040104} (\bibinfo {year} {2009})}\BibitemShut {NoStop}%
\bibitem [{\citenamefont {Langley}\ \emph {et~al.}(2018)\citenamefont
  {Langley}, \citenamefont {Lagrange}, \citenamefont {Nguyen},\ and\
  \citenamefont {Bellet}}]{Langley2018}%
  \BibitemOpen
  \bibfield  {author} {\bibinfo {author} {\bibfnamefont {D.~P.}\ \bibnamefont
  {Langley}}, \bibinfo {author} {\bibfnamefont {M.}~\bibnamefont {Lagrange}},
  \bibinfo {author} {\bibfnamefont {N.~D.}\ \bibnamefont {Nguyen}},\ and\
  \bibinfo {author} {\bibfnamefont {D.}~\bibnamefont {Bellet}},\ }\bibfield
  {title} {\bibinfo {title} {Percolation in networks of 1-dimensional objects:
  comparison between monte carlo simulations and experimental observations},\
  }\href@noop {} {\bibfield  {journal} {\bibinfo  {journal} {Nanoscale
  Horizons}\ }\textbf {\bibinfo {volume} {3}},\ \bibinfo {pages} {545}
  (\bibinfo {year} {2018})}\BibitemShut {NoStop}%
\bibitem [{\citenamefont {White}\ \emph {et~al.}(2010)\citenamefont {White},
  \citenamefont {Mutiso}, \citenamefont {Vora}, \citenamefont {Jahnke},
  \citenamefont {Hsu}, \citenamefont {Kikkawa}, \citenamefont {Li},
  \citenamefont {Fischer},\ and\ \citenamefont {Winey}}]{White2010}%
  \BibitemOpen
  \bibfield  {author} {\bibinfo {author} {\bibfnamefont {S.~I.}\ \bibnamefont
  {White}}, \bibinfo {author} {\bibfnamefont {R.~M.}\ \bibnamefont {Mutiso}},
  \bibinfo {author} {\bibfnamefont {P.~M.}\ \bibnamefont {Vora}}, \bibinfo
  {author} {\bibfnamefont {D.}~\bibnamefont {Jahnke}}, \bibinfo {author}
  {\bibfnamefont {S.}~\bibnamefont {Hsu}}, \bibinfo {author} {\bibfnamefont
  {J.~M.}\ \bibnamefont {Kikkawa}}, \bibinfo {author} {\bibfnamefont
  {J.}~\bibnamefont {Li}}, \bibinfo {author} {\bibfnamefont {J.~E.}\
  \bibnamefont {Fischer}},\ and\ \bibinfo {author} {\bibfnamefont {K.~I.}\
  \bibnamefont {Winey}},\ }\bibfield  {title} {\bibinfo {title} {Electrical
  percolation behavior in silver nanowire–polystyrene composites: Simulation
  and experiment},\ }\href {https://doi.org/10.1002/adfm.201000451} {\bibfield
  {journal} {\bibinfo  {journal} {Advanced Functional Materials}\ }\textbf
  {\bibinfo {volume} {20}},\ \bibinfo {pages} {2709} (\bibinfo {year}
  {2010})}\BibitemShut {NoStop}%
\bibitem [{\citenamefont {Mutiso}\ \emph {et~al.}(2013)\citenamefont {Mutiso},
  \citenamefont {Sherrott}, \citenamefont {Rathmell}, \citenamefont {Wiley},\
  and\ \citenamefont {Winey}}]{Mutiso2013}%
  \BibitemOpen
  \bibfield  {author} {\bibinfo {author} {\bibfnamefont {R.~M.}\ \bibnamefont
  {Mutiso}}, \bibinfo {author} {\bibfnamefont {M.~C.}\ \bibnamefont
  {Sherrott}}, \bibinfo {author} {\bibfnamefont {A.~R.}\ \bibnamefont
  {Rathmell}}, \bibinfo {author} {\bibfnamefont {B.~J.}\ \bibnamefont
  {Wiley}},\ and\ \bibinfo {author} {\bibfnamefont {K.~I.}\ \bibnamefont
  {Winey}},\ }\bibfield  {title} {\bibinfo {title} {Integrating simulations and
  experiments to predict sheet resistance and optical transmittance in nanowire
  films for transparent conductors},\ }\href
  {https://doi.org/10.1021/nn403324t} {\bibfield  {journal} {\bibinfo
  {journal} {ACS Nano}\ }\textbf {\bibinfo {volume} {7}},\ \bibinfo {pages}
  {7654} (\bibinfo {year} {2013})},\ \bibinfo {note} {pMID:
  23930701}\BibitemShut {NoStop}%
\bibitem [{\citenamefont {Loeffler}\ \emph {et~al.}(2020)\citenamefont
  {Loeffler}, \citenamefont {Zhu}, \citenamefont {Hochstetter}, \citenamefont
  {Li}, \citenamefont {Fu}, \citenamefont {Diaz-Alvarez}, \citenamefont
  {Nakayama}, \citenamefont {Shine},\ and\ \citenamefont
  {Kuncic}}]{Loeffler2020}%
  \BibitemOpen
  \bibfield  {author} {\bibinfo {author} {\bibfnamefont {A.}~\bibnamefont
  {Loeffler}}, \bibinfo {author} {\bibfnamefont {R.}~\bibnamefont {Zhu}},
  \bibinfo {author} {\bibfnamefont {J.}~\bibnamefont {Hochstetter}}, \bibinfo
  {author} {\bibfnamefont {M.}~\bibnamefont {Li}}, \bibinfo {author}
  {\bibfnamefont {K.}~\bibnamefont {Fu}}, \bibinfo {author} {\bibfnamefont
  {A.}~\bibnamefont {Diaz-Alvarez}}, \bibinfo {author} {\bibfnamefont
  {T.}~\bibnamefont {Nakayama}}, \bibinfo {author} {\bibfnamefont {J.~M.}\
  \bibnamefont {Shine}},\ and\ \bibinfo {author} {\bibfnamefont
  {Z.}~\bibnamefont {Kuncic}},\ }\bibfield  {title} {\bibinfo {title}
  {Topological properties of neuromorphic nanowire networks},\ }\href@noop {}
  {\bibfield  {journal} {\bibinfo  {journal} {Frontiers in Neuroscience}\
  }\textbf {\bibinfo {volume} {14}},\ \bibinfo {pages} {184} (\bibinfo {year}
  {2020})}\BibitemShut {NoStop}%
\bibitem [{\citenamefont {Pantone}\ \emph {et~al.}(2018)\citenamefont
  {Pantone}, \citenamefont {Kendall},\ and\ \citenamefont
  {Nino}}]{Pantone2018}%
  \BibitemOpen
  \bibfield  {author} {\bibinfo {author} {\bibfnamefont {R.~D.}\ \bibnamefont
  {Pantone}}, \bibinfo {author} {\bibfnamefont {J.~D.}\ \bibnamefont
  {Kendall}},\ and\ \bibinfo {author} {\bibfnamefont {J.~C.}\ \bibnamefont
  {Nino}},\ }\bibfield  {title} {\bibinfo {title} {Memristive nanowires exhibit
  small-world connectivity},\ }\href
  {https://doi.org/10.1016/j.neunet.2018.07.002} {\bibfield  {journal}
  {\bibinfo  {journal} {Neural Networks}\ }\textbf {\bibinfo {volume} {106}},\
  \bibinfo {pages} {144 } (\bibinfo {year} {2018})}\BibitemShut {NoStop}%
\bibitem [{\citenamefont {Watts}\ and\ \citenamefont
  {Strogatz}(1998)}]{Watts1998}%
  \BibitemOpen
  \bibfield  {author} {\bibinfo {author} {\bibfnamefont {D.~J.}\ \bibnamefont
  {Watts}}\ and\ \bibinfo {author} {\bibfnamefont {S.~H.}\ \bibnamefont
  {Strogatz}},\ }\bibfield  {title} {\bibinfo {title} {Collective dynamics of
  ‘small-world’ networks},\ }\href {https://doi.org/10.1038/30918}
  {\bibfield  {journal} {\bibinfo  {journal} {Nature}\ }\textbf {\bibinfo
  {volume} {393}},\ \bibinfo {pages} {440} (\bibinfo {year}
  {1998})}\BibitemShut {NoStop}%
\bibitem [{\citenamefont {Pascual-Garc{\'\i}a}(2016)}]{Pacual2016}%
  \BibitemOpen
  \bibfield  {author} {\bibinfo {author} {\bibfnamefont {A.}~\bibnamefont
  {Pascual-Garc{\'\i}a}},\ }\bibfield  {title} {\bibinfo {title} {A topological
  approach to the problem of emergence in complex systems},\ }\href@noop {}
  {\bibfield  {journal} {\bibinfo  {journal} {arXiv preprint arXiv:1610.02448}\
  } (\bibinfo {year} {2016})}\BibitemShut {NoStop}%
\bibitem [{\citenamefont {Strogatz}(2001)}]{Strogatz2001}%
  \BibitemOpen
  \bibfield  {author} {\bibinfo {author} {\bibfnamefont {S.~H.}\ \bibnamefont
  {Strogatz}},\ }\bibfield  {title} {\bibinfo {title} {Exploring complex
  networks},\ }\href@noop {} {\bibfield  {journal} {\bibinfo  {journal}
  {Nature}\ }\textbf {\bibinfo {volume} {410}},\ \bibinfo {pages} {268}
  (\bibinfo {year} {2001})}\BibitemShut {NoStop}%
\bibitem [{\citenamefont {Nishikawa}\ \emph {et~al.}(2003)\citenamefont
  {Nishikawa}, \citenamefont {Motter}, \citenamefont {Lai},\ and\ \citenamefont
  {Hoppensteadt}}]{Nishikawa2003}%
  \BibitemOpen
  \bibfield  {author} {\bibinfo {author} {\bibfnamefont {T.}~\bibnamefont
  {Nishikawa}}, \bibinfo {author} {\bibfnamefont {A.}~\bibnamefont {Motter}},
  \bibinfo {author} {\bibfnamefont {Y.-C.}\ \bibnamefont {Lai}},\ and\ \bibinfo
  {author} {\bibfnamefont {F.}~\bibnamefont {Hoppensteadt}},\ }\bibfield
  {title} {\bibinfo {title} {Heterogeneity in oscillator networks: Are smaller
  worlds easier to synchronize?},\ }\href@noop {} {\bibfield  {journal}
  {\bibinfo  {journal} {Physical Review Letters}\ }\textbf {\bibinfo {volume}
  {91}},\ \bibinfo {pages} {014101} (\bibinfo {year} {2003})}\BibitemShut
  {NoStop}%
\bibitem [{\citenamefont {{Jinhu Lu}}\ \emph {et~al.}(2004)\citenamefont
  {{Jinhu Lu}}, \citenamefont {{Xinghuo Yu}}, \citenamefont {{Guanrong Chen}},\
  and\ \citenamefont {{Daizhan Cheng}}}]{Lu2004}%
  \BibitemOpen
  \bibfield  {author} {\bibinfo {author} {\bibnamefont {{Jinhu Lu}}}, \bibinfo
  {author} {\bibnamefont {{Xinghuo Yu}}}, \bibinfo {author} {\bibnamefont
  {{Guanrong Chen}}},\ and\ \bibinfo {author} {\bibnamefont {{Daizhan
  Cheng}}},\ }\bibfield  {title} {\bibinfo {title} {Characterizing the
  synchronizability of small-world dynamical networks},\ }\href
  {https://doi.org/10.1109/TCSI.2004.823672} {\bibfield  {journal} {\bibinfo
  {journal} {IEEE Transactions on Circuits and Systems I: Regular Papers}\
  }\textbf {\bibinfo {volume} {51}},\ \bibinfo {pages} {787} (\bibinfo {year}
  {2004})}\BibitemShut {NoStop}%
\bibitem [{\citenamefont {Haluszczynski}\ \emph {et~al.}(2020)\citenamefont
  {Haluszczynski}, \citenamefont {Aumeier}, \citenamefont {Herteux},\ and\
  \citenamefont {Räth}}]{Haluszczynski2020}%
  \BibitemOpen
  \bibfield  {author} {\bibinfo {author} {\bibfnamefont {A.}~\bibnamefont
  {Haluszczynski}}, \bibinfo {author} {\bibfnamefont {J.}~\bibnamefont
  {Aumeier}}, \bibinfo {author} {\bibfnamefont {J.}~\bibnamefont {Herteux}},\
  and\ \bibinfo {author} {\bibfnamefont {C.}~\bibnamefont {Räth}},\ }\bibfield
   {title} {\bibinfo {title} {Reducing network size and improving prediction
  stability of reservoir computing},\ }\href
  {https://doi.org/10.1063/5.0006869} {\bibfield  {journal} {\bibinfo
  {journal} {Chaos: An Interdisciplinary Journal of Nonlinear Science}\
  }\textbf {\bibinfo {volume} {30}},\ \bibinfo {pages} {063136} (\bibinfo
  {year} {2020})}\BibitemShut {NoStop}%
\bibitem [{\citenamefont {Deng}\ and\ \citenamefont {Zhang}(2007)}]{Deng2007}%
  \BibitemOpen
  \bibfield  {author} {\bibinfo {author} {\bibfnamefont {Z.}~\bibnamefont
  {Deng}}\ and\ \bibinfo {author} {\bibfnamefont {Y.}~\bibnamefont {Zhang}},\
  }\bibfield  {title} {\bibinfo {title} {{Collective behavior of a small-world
  recurrent neural system with scale-free distribution}},\ }\href@noop {}
  {\bibfield  {journal} {\bibinfo  {journal} {IEEE Transactions on Neural
  Networks}\ }\textbf {\bibinfo {volume} {18}},\ \bibinfo {pages} {1364}
  (\bibinfo {year} {2007})}\BibitemShut {NoStop}%
\bibitem [{\citenamefont {Blondel}\ \emph {et~al.}(2008)\citenamefont
  {Blondel}, \citenamefont {Guillaume}, \citenamefont {Lambiotte},\ and\
  \citenamefont {Lefebvre}}]{Blondel2008}%
  \BibitemOpen
  \bibfield  {author} {\bibinfo {author} {\bibfnamefont {V.~D.}\ \bibnamefont
  {Blondel}}, \bibinfo {author} {\bibfnamefont {J.-L.}\ \bibnamefont
  {Guillaume}}, \bibinfo {author} {\bibfnamefont {R.}~\bibnamefont
  {Lambiotte}},\ and\ \bibinfo {author} {\bibfnamefont {E.}~\bibnamefont
  {Lefebvre}},\ }\bibfield  {title} {\bibinfo {title} {Fast unfolding of
  communities in large networks},\ }\href@noop {} {\bibfield  {journal}
  {\bibinfo  {journal} {Journal of Statistical Mechanics: Theory and
  Experiment}\ }\textbf {\bibinfo {volume} {2008}},\ \bibinfo {pages} {P10008}
  (\bibinfo {year} {2008})}\BibitemShut {NoStop}%
\bibitem [{\citenamefont {Shai}\ \emph {et~al.}(2015)\citenamefont {Shai},
  \citenamefont {Kenett}, \citenamefont {Kenett}, \citenamefont {Faust},
  \citenamefont {Dobson},\ and\ \citenamefont {Havlin}}]{Shai2015}%
  \BibitemOpen
  \bibfield  {author} {\bibinfo {author} {\bibfnamefont {S.}~\bibnamefont
  {Shai}}, \bibinfo {author} {\bibfnamefont {D.~Y.}\ \bibnamefont {Kenett}},
  \bibinfo {author} {\bibfnamefont {Y.~N.}\ \bibnamefont {Kenett}}, \bibinfo
  {author} {\bibfnamefont {M.}~\bibnamefont {Faust}}, \bibinfo {author}
  {\bibfnamefont {S.}~\bibnamefont {Dobson}},\ and\ \bibinfo {author}
  {\bibfnamefont {S.}~\bibnamefont {Havlin}},\ }\bibfield  {title} {\bibinfo
  {title} {Critical tipping point distinguishing two types of transitions in
  modular network structures},\ }\href
  {https://doi.org/10.1103/PhysRevE.92.062805} {\bibfield  {journal} {\bibinfo
  {journal} {Phys. Rev. E}\ }\textbf {\bibinfo {volume} {92}},\ \bibinfo
  {pages} {062805} (\bibinfo {year} {2015})}\BibitemShut {NoStop}%
\bibitem [{\citenamefont {Newman}(2000)}]{Newman2000}%
  \BibitemOpen
  \bibfield  {author} {\bibinfo {author} {\bibfnamefont {M.}~\bibnamefont
  {Newman}},\ }\bibfield  {title} {\bibinfo {title} {Models of the small
  world},\ }\href@noop {} {\bibfield  {journal} {\bibinfo  {journal} {Journal
  of Statistical Physics}\ }\textbf {\bibinfo {volume} {101}},\ \bibinfo
  {pages} {819} (\bibinfo {year} {2000})}\BibitemShut {NoStop}%
\bibitem [{\citenamefont {Muldoon}\ \emph {et~al.}(2016)\citenamefont
  {Muldoon}, \citenamefont {Bridgeford},\ and\ \citenamefont
  {Bassett}}]{Muldoon2016}%
  \BibitemOpen
  \bibfield  {author} {\bibinfo {author} {\bibfnamefont {S.~F.}\ \bibnamefont
  {Muldoon}}, \bibinfo {author} {\bibfnamefont {E.~W.}\ \bibnamefont
  {Bridgeford}},\ and\ \bibinfo {author} {\bibfnamefont {D.~S.}\ \bibnamefont
  {Bassett}},\ }\bibfield  {title} {\bibinfo {title} {Small-world propensity
  and weighted brain networks},\ }\href {https://doi.org/10.1038/srep22057}
  {\bibfield  {journal} {\bibinfo  {journal} {Scientific Reports}\ }\textbf
  {\bibinfo {volume} {6}},\ \bibinfo {pages} {22057} (\bibinfo {year}
  {2016})}\BibitemShut {NoStop}%
\bibitem [{\citenamefont {Barahona}\ and\ \citenamefont
  {Pecora}(2002)}]{Barahona2002}%
  \BibitemOpen
  \bibfield  {author} {\bibinfo {author} {\bibfnamefont {M.}~\bibnamefont
  {Barahona}}\ and\ \bibinfo {author} {\bibfnamefont {L.~M.}\ \bibnamefont
  {Pecora}},\ }\bibfield  {title} {\bibinfo {title} {Synchronization in
  small-world systems},\ }\href {https://doi.org/10.1103/PhysRevLett.89.054101}
  {\bibfield  {journal} {\bibinfo  {journal} {Phys. Rev. Lett.}\ }\textbf
  {\bibinfo {volume} {89}},\ \bibinfo {pages} {054101} (\bibinfo {year}
  {2002})}\BibitemShut {NoStop}%
\bibitem [{\citenamefont {Hagberg}\ \emph {et~al.}(2008)\citenamefont
  {Hagberg}, \citenamefont {Schult},\ and\ \citenamefont {Swart}}]{networkx}%
  \BibitemOpen
  \bibfield  {author} {\bibinfo {author} {\bibfnamefont {A.~A.}\ \bibnamefont
  {Hagberg}}, \bibinfo {author} {\bibfnamefont {D.~A.}\ \bibnamefont
  {Schult}},\ and\ \bibinfo {author} {\bibfnamefont {P.~J.}\ \bibnamefont
  {Swart}},\ }\bibfield  {title} {\bibinfo {title} {Exploring network
  structure, dynamics, and function using networkx},\ }in\ \href@noop {} {\emph
  {\bibinfo {booktitle} {Proceedings of the 7th Python in Science
  Conference}}},\ \bibinfo {editor} {edited by\ \bibinfo {editor}
  {\bibfnamefont {G.}~\bibnamefont {Varoquaux}}, \bibinfo {editor}
  {\bibfnamefont {T.}~\bibnamefont {Vaught}},\ and\ \bibinfo {editor}
  {\bibfnamefont {J.}~\bibnamefont {Millman}}}\ (\bibinfo {address} {Pasadena,
  CA USA},\ \bibinfo {year} {2008})\ pp.\ \bibinfo {pages} {11 --
  15}\BibitemShut {NoStop}%
\bibitem [{\citenamefont {Csardi}\ and\ \citenamefont {Nepusz}(2006)}]{igraph}%
  \BibitemOpen
  \bibfield  {author} {\bibinfo {author} {\bibfnamefont {G.}~\bibnamefont
  {Csardi}}\ and\ \bibinfo {author} {\bibfnamefont {T.}~\bibnamefont
  {Nepusz}},\ }\bibfield  {title} {\bibinfo {title} {The igraph software
  package for complex network research},\ }\href@noop {} {\bibfield  {journal}
  {\bibinfo  {journal} {InterJournal}\ }\textbf {\bibinfo {volume} {Complex
  Systems}},\ \bibinfo {pages} {1695} (\bibinfo {year} {2006})}\BibitemShut
  {NoStop}%
\bibitem [{\citenamefont {Rubinov}\ and\ \citenamefont
  {Sporns}(2010)}]{matlab}%
  \BibitemOpen
  \bibfield  {author} {\bibinfo {author} {\bibfnamefont {M.}~\bibnamefont
  {Rubinov}}\ and\ \bibinfo {author} {\bibfnamefont {O.}~\bibnamefont
  {Sporns}},\ }\bibfield  {title} {\bibinfo {title} {Complex network measures
  of brain connectivity: Uses and interpretations},\ }\href
  {https://doi.org/10.1016/j.neuroimage.2009.10.003} {\bibfield  {journal}
  {\bibinfo  {journal} {NeuroImage}\ }\textbf {\bibinfo {volume} {52}},\
  \bibinfo {pages} {1059 } (\bibinfo {year} {2010})},\ \bibinfo {note}
  {computational Models of the Brain}\BibitemShut {NoStop}%
\bibitem [{\citenamefont {Clauset}\ \emph {et~al.}(2004)\citenamefont
  {Clauset}, \citenamefont {Newman},\ and\ \citenamefont
  {Moore}}]{Clauset2004}%
  \BibitemOpen
  \bibfield  {author} {\bibinfo {author} {\bibfnamefont {A.}~\bibnamefont
  {Clauset}}, \bibinfo {author} {\bibfnamefont {M.~E.~J.}\ \bibnamefont
  {Newman}},\ and\ \bibinfo {author} {\bibfnamefont {C.}~\bibnamefont
  {Moore}},\ }\bibfield  {title} {\bibinfo {title} {Finding community structure
  in very large networks},\ }\href {https://doi.org/10.1103/PhysRevE.70.066111}
  {\bibfield  {journal} {\bibinfo  {journal} {Phys. Rev. E}\ }\textbf {\bibinfo
  {volume} {70}},\ \bibinfo {pages} {066111} (\bibinfo {year}
  {2004})}\BibitemShut {NoStop}%
\bibitem [{\citenamefont {Newman}(2006)}]{Newman2006}%
  \BibitemOpen
  \bibfield  {author} {\bibinfo {author} {\bibfnamefont {M.~E.~J.}\
  \bibnamefont {Newman}},\ }\bibfield  {title} {\bibinfo {title} {Modularity
  and community structure in networks},\ }\href
  {https://doi.org/10.1073/pnas.0601602103} {\bibfield  {journal} {\bibinfo
  {journal} {Proceedings of the National Academy of Sciences}\ }\textbf
  {\bibinfo {volume} {103}},\ \bibinfo {pages} {8577} (\bibinfo {year}
  {2006})}\BibitemShut {NoStop}%
\end{thebibliography}%

\end{document}